\documentclass[journal]{aa}

\usepackage{graphicx}
\usepackage{amsmath}
\usepackage{amssymb}
\usepackage{verbatim}
\usepackage{array}
\usepackage{txfonts}
\usepackage{natbib}

\bibpunct{(}{)}{;}{a}{}{,}

\newcommand{\celltspace}{\rule{0pt}{2.8ex}}
\newcommand{\cellbspace}{\rule[-1.4ex]{0pt}{0pt}}

\defcitealias{Chan:1993}{CL93}

\begin{document}

\title{Annihilation emission from young supernova remnants}
\titlerunning{Annihilation emission from young supernova remnants}
\author{Pierrick Martin \inst{1} \and Jacco Vink \inst{2} \and Sarka Jiraskova \inst{2,3} \and Pierre Jean \inst{4,5} \and Roland Diehl \inst{1}}
\institute{Max Planck Institut f\"ur extraterrestrische Physik, Postfach 1312, 85741 Garching, Germany 
              \and  Astronomical Institute, University Utrecht, PO Box 80000, 3508 TA Utrecht, The Netherlands
              \and Department of Astrophysics/IMAPP, Radboud University Nijmegen, PO Box 9010, 6500 GL Nijmegen, The Netherlands
              \and Centre d'Etude Spatiale des Rayonnements, CNRS/UPS, 9, avenue colonel Roche, BP44346, 31028 Toulouse cedex 4, France 
              \and Universit\'e de Toulouse (UPS), Centre National de la Recherche Scientifique (CNRS), UMR 5187}
\date{Received: 01 February 2010 / Accepted: 02 June 2010}
\abstract{A promising source of the positrons that contribute through annihilation to the diffuse Galactic 511\,keV emission is the \mbox{$\beta^+$-decay} of unstable nuclei like $^{56}$Ni and $^{44}$Ti synthesised by massive stars and supernovae. Although a large fraction of these positrons annihilate in the ejecta of SNe/SNRs, no point-source of annihilation radiation appears in the INTEGRAL/SPI map of the 511\,keV emission.}{We exploit the absence of detectable annihilation emission from young local SNe/SNRs to derive constraints on the transport of MeV positrons inside SN/SNR ejecta and their escape into the CSM/ISM, both aspects being crucial to the understanding of the observed Galactic 511\,keV emission.}{We simulated 511\,keV lightcurves resulting from the annihilation of the decay positrons of $^{56}$Ni and $^{44}$Ti in SNe/SNRs and their surroundings using a simple model. We computed specific 511\,keV lightcurves for Cas A, Tycho, Kepler, SN1006, G1.9+0.3 and SN1987A, and compared these to the upper-limits derived from INTEGRAL/SPI observations.}{The predicted 511\,keV signals from positrons annihilating in the ejecta are below the sensitivity of the SPI instrument by several orders of magnitude, but the predicted 511\,keV signals for positrons escaping the ejecta and annihilating in the surrounding medium allowed to derive upper-limits on the positron escape fraction of $\sim$13\% for Cas A, $\sim$12\% for Tycho, $\sim$30\% for Kepler and $\sim$33\% for SN1006.}{The transport of $\sim$MeV positrons inside SNe/SNRs cannot be constrained from current observations of the 511\,keV emission from these objects, but the limits obtained on their escape fraction are consistent with a nucleosynthesis origin of the positrons that give rise to the diffuse Galactic 511\,keV emission.}

\maketitle

\section{Introduction}
\label{intro}

\indent One of the most valuable contributions of the INTEGRAL space-borne gamma-ray observatory so far is the cartography of the 511\,keV galactic emission from electron-positron annihilation \citep{Knodlseder:2005,Weidenspointner:2008}. This achievement was obtained from several years of observation with the high-resolution SPI spectrometer and provides us with the most accurate view on leptonic antimatter in the Galaxy. This result, and its expected improvement over the coming years as exposure grows (INTEGRAL is planned to operate till at least 2012), will very likely help clarifying the origin of the positrons that annihilate in our Galaxy, an issue that has been lasting for several decades and that has proven over the years to be of interest for a number of different science fields, from astrophysics to particle and plasma physics.\\
\indent A review of the history and stakes of annihilation gamma-ray observations, from early balloon experiments to the SMM/GRS and CGRO/OSSE satellites, can be found in \citet{Prantzos:2010} and \citet{Diehl:2009}. The above-mentioned INTEGRAL result confirms the basic description of the emission that followed from the CGRO mission, a bright bulge typically 10$^{\circ}$ in width on top of a faint disk. The positive latitude enhancement found from OSSE data is however not observed by SPI and instead, a longitudinal asymmetry in the inner disk is found \citep{Weidenspointner:2008,Weidenspointner:2008a}. The spectrometric capabilities of the SPI instrument also allowed to derive a high-resolution annihilation spectrum, and the observed line profile indicates that most of the annihilation occurs through positronium formation and takes place in the warm phases of the interstellar medium \citep{Jean:2006}.\\
\indent The observed intensity distribution, with a high inferred bulge-to-disk ratio of 2-6 in luminosity \citep{Weidenspointner:2008a}, is intriguing in that it does not correlate with any distribution of classical sources or potential annihilation medium. For instance, thermonuclear supernovae (SNIa), that are promising positron sources, are $\sim$ 10 times more numerous in the disk than in the bulge and would give a bulge-to-disk luminosity ratio of $\sim$ 0.1 if the positrons were to annihilate not too far from their sources \citep{Prantzos:2006}. The peculiar morphology of the galactic 511\,keV emission has been invoked as an evidence for an exotic origin of the positrons, such as dark matter decay in the inner regions of the Galaxy \citep{Bohm:2004a}. Another explanation put forward is the large-scale transport of positrons created by more classical sources \citep{Prantzos:2006,Higdon:2009}. The latter alternative is supported to some extent by the spectroscopic properties of the annihilation line. Indeed, most plausible positron sources are found in the hot phase, which occupies most of the interstellar volume, while most positrons are observed to annihilate in the warm phase \citep{Jean:2006,Churazov:2005}; positron transport therefore seems to be required to a certain degree. Overall, the discussions raised by the INTEGRAL observations illustrate that understanding the observed annihilation emission is both a problem of sources and transport of the positrons.\\
\indent Among the most plausible sources of galactic positrons is the $\beta^+$ decay of unstable nuclei synthesised by massive stars and supernovae, mainly $^{56}$Ni, $^{44}$Ti and $^{26}$Al. The existence and decay of the above-mentioned isotopes are directly established by gamma-ray line observations \citep{Leising:1990,Renaud:2006,Martin:2009}, and the observed or inferred isotope yields are fairly sufficient to account for the steady state annihilation rate of a few 10$^{43}$\,e$^{+}$\,s$^{-1}$ deduced from INTEGRAL observations. Under the assumption that positrons annihilate close to their sources, the estimated annihilation rate in the disk of the Galaxy agrees with the estimated positron production rate from $^{26}$Al decay \citep{Knodlseder:2005}. Because of the high bulge-to-disk ratio in luminosity, however, the origin of the positrons annihilating in the bulge of the Galaxy is much more uncertain. From COMPTEL observations, the initial energy of these positrons was constrained to be below a few MeV, otherwise they would give rise to a diffuse MeV emission of in-flight annihilation\footnote{Very energetic positrons can also give rise to a GeV Galactic diffuse emission through inverse-Compton and bremsstrahlung radiation, as do the $\sim$ 10 times less numerous secondary positrons coming from hadronic interactions of cosmic-rays with ambient gas \citep[see][]{Strong:2004,Aharonian:2000}.} \citep[see for instance][]{Beacom:2006,Sizun:2006}. In that respect, the positrons from nucleosynthesis are good candidates since they are released with initial kinetic energies of $\sim$ 1\,MeV \citep[but see][]{Chernyshov:2010}.\\
\indent In order to contribute to the 511\,keV galactic annihilation emission, the decay positrons must first be slowed down to below $\sim$ 100\,eV and then annihilate in optically thin environments. Since the interstellar medium is fully transparent to 511\,keV radiation, this means that the parent radio-isotopes and/or their positrons should escape the dense stellar interiors where they have been produced. The slowing-down and annihilation of decay positrons may take a certain time (see \ref{decaypositrons_life}), during which the initially relativistic particles can diffuse. The ultimate fate of decay positrons therefore depends on the characteristics of their transport, first in the expanding stellar ejecta of supernovae (SNe) and supernova remnants (SNRs) and then in the circumstellar medium (CSM) and interstellar medium (ISM). The transport of MeV positrons, however, is as yet largely unsolved. The nature of their diffusion regime, collisionless (positrons efficiently scatter off MHD waves) or collisional (positrons follow unperturbed trajectories along magnetic field lines\footnote{The term "unperturbed" refers here to the absence of efficient scattering, but the particle constantly loses energy along its trajectory. Coulomb or ionisation/excitation interactions involve relatively small energy variations for MeV positrons and act like continuous energy-loss processes with small pitch-angle scattering. The helicoidal motion of the particle around a magnetic field line is thus only marginally affected.}), is still unknown. This can affect the characteristic path lengths travelled by positrons before annihilation by several orders of magnitude, from several pc to a few ten kpc. The problem stems firstly from our incomplete understanding of how MHD turbulence scatters particles through their interaction with MHD waves. Secondly, we have a poor knowledge of the MHD turbulence at small spatial scales (of the order of the Larmor radius of MeV positrons in $\sim \mu$G fields), where the decay positrons could resonantly interact with perturbation modes. Recent works indicate that the cascade of MHD fluctuations is quenched at too large spatial scales as a result of ion-neutral collisions \citep{Higdon:2009,Jean:2009}. In this case, no collisionless diffusion through resonant interactions takes place and positrons can travel over large linear distances. Yet, many aspects of the issue should still be investigated, like non-resonant interactions with fast magnetosonic modes, which could lead to efficient diffusion even if no magnetic fluctuations exist at the resonant scale, or perpendicular diffusion by field line wandering \citep{Prantzos:2010}. A complete description of positron transport will also need to take into account diffusion in momentum space (reacceleration). On the whole, the issue is far from settled and theoretical as well as observational efforts are still needed.\\
\indent A nucleosynthesis origin for the Galactic positrons therefore requires a sufficient fraction of decay positrons to escape the production sites of their parent isotopes and feed the galactic population, and a global transport of these particles in the ISM so that the distribution of their annihilation sites can account for the observed morphology.\\
\indent As we will see later, a large fraction of the positrons created in the decay of $^{56}$Ni and $^{44}$Ti are thought to annihilate inside SNe and SNRs. Despite this fact, no point-like emission appears in the INTEGRAL 511\,keV map. This non-detection may therefore be translated into some evidence concerning the intertwined aspects of MeV positron transport and escape, both topics being crucial to the understanding of the observed Galactic annihilation emission.

\section{Decay positrons in SNe and SNRs}
\label{decaypositrons}

\indent Positron production by the three above-mentioned radio-isotopes occurs in the following way:
\begin{align}
& ^{56}\textrm{Ni} \, \xrightarrow{9\,\textrm{d} \,} \, ^{56}\textrm{Co} \, \xrightarrow{111\,\textrm{d} \,} \, ^{56}\textrm{Fe} + \textrm{e}^+ \quad (18.1\%) \notag \\
& ^{44}\textrm{Ti} \, \xrightarrow{85\,\textrm{yr} \,} \, ^{44}\textrm{Sc} \, \xrightarrow{6\,\textrm{h} \,} \, ^{44}\textrm{Ca} + \textrm{e}^+ \quad (94.3\%) \notag \\
& ^{26}\textrm{Al} \, \xrightarrow{1\,\textrm{Myr} \,} \, ^{26}\textrm{Mg} + \textrm{e}^+ \quad (81.7\%) \notag 
\end{align}
where the decay timescales and $\beta$-fractions are given for each decay chain. The electron neutrinos and antineutrinos (required for conservation of leptonic number) and the gamma photons (from the de-excitation of daughter nuclei) have been omitted. The positrons from $^{56}$Ni and $^{44}$Ti are actually released in the decay of their daughter nuclei, but in the following we will keep the term $^{56}$Ni and $^{44}$Ti positrons to emphasise the connection with nucleosynthesis products.\\
\indent In the context of SNe/SNRs, the specificities of these most likely contributors to the Galactic positron budget are the following:
\begin{enumerate}
\item $^{56}$Ni is synthesised by explosive Si-burning deep in the stellar ejecta during core-collapse and thermonuclear SN explosions, with typical yields of a few 10$^{-1}$\,M$_{\odot}$. It is short-lived and escape is therefore a quite critical problem as the dense medium of the late SN/early SNR is likely to stop all positrons and absorb some of the annihilation radiation.
\item $^{44}$Ti is synthesised by explosive Si-burning deep in the stellar ejecta during core-collapse and thermonuclear SN explosions, with typical yields of a few 10$^{-4}$-10$^{-5}$\,M$_{\odot}$. It is medium-lived and depending on the conditions, the escape fraction can vary between two extremes (no or total escape). Yet, the diluted ejecta at typical decay time is mostly transparent to gamma-rays and so an annihilation signal can be expected from those positrons that cannot escape the stellar envelope.
\item $^{26}$Al is synthesised at various stages of the evolution of a massive star with typical yields of a few 10$^{-4}$-10$^{-5}$\,M$_{\odot}$. It is ejected in the ISM through stellar winds and SN explosions. Since it is long-lived, escape is not a problem and all decay positrons are directly injected in the ISM.
\end{enumerate}
In the present work, we are interested in point-sources of annihilation radiation so we will not consider $^{26}$Al positrons, which are expected to give rise to a diffuse annihilation signal (in the Galactic disk in particular). We therefore focus on $^{56}$Ni and $^{44}$Ti positrons only in the context of SNe and SNRs. Before addressing the specific case of positron transport and annihilation in SNe/SNRs, however, we describe in more detail the main physical processes that rule the life of a MeV positron until its annihilation.

\subsection{Life and death of MeV positrons}
\label{decaypositrons_life}

\indent Once created in the $\beta$-decay of a given radionucleus, positrons are slowed down by ionisation/excitation and Coulomb losses. If their kinetic energy is brought down to $\sim$ 100\,eV, they can rip electrons off atoms (a process called charge exchange) and form positronium, which is a short-lived electron-positron bound state (the overall process is termed "positronium formation in-flight"). Depending on spin orientation, positronium then quickly decays in two 511\,keV photons (25\% of cases) or three continuum photons (75\% of cases, with the total radiated energy being equal to $2\,m_{e}c^{2}$ in the positronium rest frame). The positrons that do not form positronium in-flight eventually thermalise and annihilate directly with bound or free electrons, thereby giving two 511\,keV photons, or through positronium formation by radiative recombination with free electrons \citep[for a review of the various annihilation channels, see][]{Guessoum:2005}. It should be noted here that the cross-sections for positronium formation by charge exchange are several orders of magnitude larger than for any other process. So unless the medium is strongly ionised (and hence charge exchange cannot efficiently take place), most positrons will form positronium and then quickly annihilate once they are slowed down to a kinetic energy of $\sim$ 100\,eV. If the medium is strongly ionised, positronium formation by radiative recombination of thermalised particles is the dominant process, provided the temperature is moderate (for temperatures above $\sim$ 10$^{6}$\,K, the Maxwellian distribution of particles is shifted to an energy range where direct annihilation with free electrons takes over).\\
\indent The physical properties of the medium therefore influence the channel by which most positrons annihilate, and hence the fraction of positrons that annihilate through 511\,keV emission as opposed to continuum emission. Different channels also imply different lifetimes for the positrons. The typical time to slow down a MeV positron down to $\sim$ 100\,eV through ionisation/excitation or Coulomb losses is $\sim 10^{4-5}$ $n^{-1}$\,yrs (where $n$ is the medium density in cm$^{-3}$) and the time for complete thermalisation is similar since the energy-loss rate strongly increases as the particle energy decreases. Then, the time for positronium formation by charge exchange is comparatively negligible so for this annihilation channel, the lifetime of positrons is imposed by the slowing-down time. In contrast, the time for direct annihilation of thermalised positrons with free or bound electrons is $\sim 10^{5-6}$ $n^{-1}$\,yrs, so in the latter scenario the lifetime of positrons is imposed by the annihilation time. In all cases, however, the range of positrons is imposed by the slowing-down period.

\subsection{MeV positrons in SNe/SNRs}
\label{decaypositrons_sne}

\indent A SN ejecta is a relatively dense medium in which the decay positrons will be slowed down quite efficiently in early times and then on longer and longer timescales as the stellar material gets diluted. For the simple case of a uniform ejecta in homologous expansion and under the assumption that positrons remain confined to the stellar ejecta and do not escape, the slowing-down time for a MeV positron injected at a given time after the explosion can be approximated\footnote{The formula gives the slowing-down time assuming that the positron injected at a given time experiences the same ejecta density during its slowing-down. As time goes by and slowing-down time increases, this assumption becomes less and less valid and the estimated value turns to a lower-limit.} by:
\begin{equation}
t_{sd} \simeq 10^{-3} \times \frac{E_{ej}^{3/2}}{M_{ej}^{5/2}} \times t^{3} \quad \textrm{yrs}
\end{equation}
where $M_{ej}$ and $E_{ej}$ are the ejecta mass and energy respectively (in M$_{\odot}$ and 10$^{51}$\,erg) and $t$ the time since the explosion at which the positron is injected (in yrs).\\
\indent Owing to its very rapid expansion, the stellar ejecta is thought to be quite cold and hence neutral, which means that positrons are slowed down by ionisation/excitation losses and then annihilate mostly through positronium formation in-flight just after the slowing-down period. From the above formula, we expect the slowing-down time to be below a year during the first years after the SN explosion. Over this period, the annihilation lightcurve therefore follows the radioactive decay curve. Then, between 10 and 100\,yrs, the slowing-down time increases up to few centuries and so the positrons created in this interval and trapped in the ejecta will experience a delayed annihilation over a characteristic timescale that corresponds to the typical ages of the youngest Galactic SNRs (a few centuries to a millenary). After a century, the slowing-down times become too long for annihilation to occur inside the SNR and the corresponding positrons very likely escape. For the simple case we are considering here, we therefore expect most positrons from the short-lived $^{56}$Ni to annihilate soon after their release in the ejecta. Yet, a tiny fraction of them will give rise to a delayed annihilation signal which, given the high yields of $^{56}$Ni, may be in reach of modern gamma-ray instruments. $^{44}$Ti holds more potential for emission from SNRs because of its $\sim$ 100\,yr lifetime. At early times, positrons annihilate nearly instantly, as in the case of $^{56}$Ni, but the sizeable fraction of positrons created after a few decades will lead to a delayed annihilation, potentially observable today in the youngest Galactic SNRs.\\
\indent This scenario is rather simplified and many aspects of the problem could alter the above conclusions. First of all, the above results are valid for 1\,MeV positrons whereas most decay positrons actually have lower energies of a few 100\,keV. Then, we made the assumption that decay positrons are "trapped" in the ejecta, which appears as the most favourable case for efficient slowing-down and subsequent annihilation, but positron escape very likely occurs (and actually should occur if the Galactic annihilation emission, especially from the Galactic bulge, is to be traced to a nucleosynthesis origin), which would lead to reduced 511\,keV fluxes. On the other hand, escaping positrons that lost some fraction of their kinetic energy during ejecta crossing can achieve their slowing-down and annihilate in the circumstellar environment of the SNR.\\
\indent Our order-of-magnitude evaluation, together with the above considerations, show that there is potential for detectable annihilation radiation from SNe/SNRs up to a few centuries after the explosion, but such an emission is not observed. INTEGRAL/SPI observations provide the most stringent constraints on annihilation in SNe and SNRs so far, and therefore call for a deeper interpretation of the results. In this work, we focused on the 511\,keV emission from a few young and well-characterized SNRs to constrain the transport of positrons in the ejecta and in the CSM/ISM. Upper-limits on the 511\,keV annihilation flux from six young SNRs, Cassiopeia A, Tycho, Kepler SN1006, SN1987A and G1.9+0.3, were derived from INTEGRAL/SPI data for various annihilation scenarios. These constraints were then compared to simulated annihilation lightcurves computed in the frame of simple descriptions for the SN/SNR evolution and CSM/ISM environment, and under various prescriptions for the transport of the particles.

\section{Past works}
\label{pastworks}

\indent In order to evaluate the contribution of decay positrons to the galactic annihilation radiation observed by the GRIS and OSSE instruments, \citet{Chan:1993} performed a theoretical study of the escape of $^{56}$Ni and $^{44}$Ti positrons from SNe of both types (thermonuclear SNIa and core-collapse SNe). Their results indicate that between 0.1 and 15\% of $^{56}$Ni positrons can escape in deflagration or delayed detonation models of SNIa, as well as in core-collapse SNe models. In contrast, the escape fractions of $^{44}$Ti positrons range from 30 to 100\%. The positrons that survived then enter the ISM where their slowing-down will proceed on longer timescales. The typical lifetime of MeV positrons in the ISM is about 10$^{4}$-10$^{8}$\,yr, depending on the ISM phase \citep{Jean:2006}, which allows their population to build up from many successive SNe. From their results on $^{56}$Ni and $^{44}$Ti, supplemented by an observational estimate of the positrons from $^{26}$Al, \citet{Chan:1993} showed that a steady-state production (and annihilation) rate of a few 10$^{43}$\,e$^{+}$\,s$^{-1}$ could be achieved, in agreement with the (model-dependent) value derived from observations. The study also illustrated the critical role played by the magnetic field topology inside the ejecta (which could confine positrons in the ejecta if it is strong and turbulent enough, thereby favouring energy losses and annihilation), and to a lesser extent by ejecta mixing (which could lift iron to the outermost lowest density regions of the ejecta, thereby favouring positron escape) and explosion energy (which controls the rate at which the ejecta becomes transparent to positrons). For instance, with the standard W7 deflagration model of \citet{Nomoto:1984}, the authors found that the escape fraction of $^{56}$Ni positrons in a SNIa ranges from 0.1 to 5\% depending on whether the magnetic field is thorougly tangled or fully combed out, and this increases to 2.5 to 13\% when the ejecta is uniformly mixed.\\
\indent Another way of estimating the escape fraction of decay positrons is based on the study of SNIa late lightcurves. The bolometric lightcurve of a SNIa is indeed powered by $^{56}$Ni positron energy deposition after the ejecta became optically thin to the $^{56}$Ni gamma decay photons, typically for t $\geq$ 200\,days. The driving parameters of the problem are again the structure and strength of the magnetic field inside the ejecta but also the ionisation state of the ejecta because an ionised medium is more efficient at slowing charged particles than is a neutral medium. Depending on whether positrons are trapped locally in the ejecta by a strong turbulent magnetic field or whether they can flow out almost freely along a weak radial magnetic field, the simulated bolometric lightcurves are different. In particular, deviations from one or the other scenario are detectable at late times, between 400 and 1000\,days after explosion. From a set of models, \citet{Milne:1999} derived positron escape fractions in the range $\sim$0 to 11\%, in agreement with the values found by \citet{Chan:1993} (who did not include ionisation effects in their study). From a comparison with a sample of 10 lightcurves from a variety of SNe Ia (derived mostly from B- and V-band photometry), the authors concluded that positron escape occurs in SNIa ejecta and that a radial or weak magnetic field configuration is indicated by the observations. For the standard W7 deflagration model and the assumption of a radial magnetic field, they computed an escape fraction in the range 1.8 to 5.5\%, depending on the ionisation stage of the ejecta, with the higher value being favoured by the observations. Yet, \citet{Lair:2006} showed from BVRI photometry that current data on late lightcurves of SNIa are insufficient to assess if color evolution and shift of the luminosity to longer wavelengths occur in the SNe Ia. Consequently, late bolometric light-curves extrapolated from optical bands could be increasingly inaccurate with time and would thus jeopardise the evaluation of the positron escape fraction.\\
\indent A third way to determine the escape fraction was explored by \citet{Kalemci:2006} and is based on the annihilation emission from SN1006, the remnant of a SNIa. The authors first used the INTEGRAL/SPI data to obtain an upper limit on the 511keV flux from this object. Then, under the assumptions that escaped positrons can be locally confined close to the SNR and that their mean lifetime is less than 10$^{5}$\,yrs, they derive an upper limit on the escape fraction of 7.5\%. These two strong assumptions on the diffusion and lifetime of positrons in the ISM (or CSM) are, however, not supported by any observational or even theoretical argument. We will thereafter present a more detailed treatment of the annihilation of $^{56}$Ni positrons in the ISM (and in the ejecta), which will allow a more accurate interpretation of the upper-limit on the 511\,keV flux from SN1006.\\
\indent Most of the above-mentioned studies of positrons from SNe and SNRs were mainly focused on evaluating if these particles can escape and then fill the ISM and eventually feed the galactic annihilation emission. Yet, the direct annihilation emission from SNe and SNRs has been poorly studied. The simulations of SNIa late lightcurves including the contribution from decay positrons could have been the opportunity to provide simulated annihilation lightcurves from SNe Ia and their early remnants, but the limited time range of these simulations (up to $\sim$ 1000\,days) and the absence of constraints from gamma-ray line observations probably prevented the effort to be continued in that direction. INTEGRAL/SPI opened the way for cartography of the annihilation emission of diffuse or point-like origin. In \citet{Knodlseder:2005}, the authors exploited these capabilities to analyse the 511\,keV emission from the Galaxy, including its extended component and the possible contributions from a multitude of point-like objects like SNRs. By now, the volume of available SPI data is as much as 6 times larger than the data set used by \citet{Knodlseder:2005}. Conditions are therefore favourable for a new search for annihilation emission from SNRs.
\begin{table*}[!t]
\begin{minipage}[][10cm][c]{\textwidth}
\begin{center}
\caption{Upper limits on the 511\,keV flux in units of 10$^{-5}$\,ph\,cm$^{-2}$\,s$^{-1}$, in three different energy bands and for various source sizes for SN1006 and Tycho. Also listed are the INTEGRAL/SPI exposure for each object in cm$^{2}$\,s.}
\begin{tabular}{|c|c|c|c|c|c|c|}
\hline
\celltspace SNR & SPI exposure & Sky model & 510-512\,keV & 508.5-513.5\,keV & 506-516\,keV \cellbspace \\
\hline
\celltspace Cassiopeia A & $\sim 3 \times 10^{8}$ & point-source & 2.8 & 3.9 & 4.7 \cellbspace \\
\hline
\celltspace Tycho & $\sim 3 \times 10^{8}$ & point-source & 2.6 & 3.8 & 4.6 \cellbspace \\
\celltspace            &  & 3$^{\circ}$ gaussian &  & 5.4 &  \cellbspace \\
\celltspace            &  & 6$^{\circ}$ gaussian &  & 7.5 &  \cellbspace \\
\hline
\celltspace Kepler & $\sim 8 \times 10^{8}$ & point-source & 1.8 & 2.5 & 3.0 \cellbspace \\
\hline
\celltspace SN1006 & $\sim 3 \times 10^{8}$ & point-source & 2.8 & 4.0 & 4.8 \cellbspace \\
\celltspace               &  & 3$^{\circ}$ gaussian &  & 5.1 &  \cellbspace \\
\celltspace               &  & 6$^{\circ}$ gaussian &  & 6.4 &  \cellbspace \\
\celltspace               &  & 9$^{\circ}$ gaussian &  & 7.2 &  \cellbspace \\
\celltspace               &  & 12$^{\circ}$ gaussian &  & 7.9 &  \cellbspace \\
\celltspace               &  & 15$^{\circ}$ gaussian &  & 8.5 &  \cellbspace \\
\hline
\celltspace SN1987A & $\sim 8 \times 10^{7}$ & point-source & 3.8 & 5.4 & 6.6 \cellbspace \\
\hline
\celltspace G1.9+0.3 & $\sim 9 \times 10^{8}$ & point-source & 1.8 & 2.6 & 3.1 \cellbspace \\
\hline
\end{tabular}
\label{tab_obs}
\end{center}
\end{minipage}
\end{table*}

\section{INTEGRAL/SPI observations}
\label{data}

\subsection{Instrument characteristics}
\label{data_instru}

The SPI instrument onboard the INTEGRAL gamma-ray space observatory is a high-resolution spectrometer with imaging capabilities \citep{Vedrenne:2003,Roques:2003}. The spectrometry of gamma radiations with energies between 20 keV and 8 MeV is performed by an array of 19 high-purity germanium detectors (as yet 17, since detector 2 and 17 failed on revolution 142 and 210 respectively), with a spectral resolution of about 2 keV FWHM at 511\,keV. All gamma-ray sources, diffuse or point-like, emitting in this energy range can also be imaged indirectly thanks to a coded mask system with an angular resolution of 2.8$^{\circ}$ and a fully-coded field of view of 16$^{\circ}$x16$^{\circ}$.\\
\indent The data produced by SPI are strongly background-dominated (signal-to-noise ratio of less than 1\%) because substantial hadronic interactions and electromagnetic cascades are induced by solar and cosmic-ray particles impacting the satellite. These result in an instrumental background composed of a continuum and a certain number of lines of various intensities. In particular, the 511\,keV line appears in the background spectrum and is even one of the strongest lines. Extracting the weak gamma-ray line signals of astrophysical origin therefore requires first an accurate modelling of that background noise. Then, the reconstruction of a given source intensity distribution is achieved through multiple exposures (following a \textit{dithering pattern}) and model-fitting approaches (as explained below).

\subsection{Data set}
\label{data_set}

The data used for this analysis are all-sky data that have been collected between revolution 7 and 764 of the INTEGRAL satellite (the duration of one INTEGRAL orbit is $\sim$3 days). A filtering has been applied to exclude all pointings during which abnormally high count-rates (due to solar flares and periodic radiation belt crossings) occurred and consequently preclude the detection of celestial signals. The total effective observation time eventually amounts to 91.8\,Ms, among which 58.6\,Ms cover the galactic plane ($\lvert$b$\rvert$ $\leq$ 20$^{\circ}$) and 33.2\,Ms are high-latitude pointings ($\lvert$b$\rvert$ $>$ 20$^{\circ}$). All \textit{single events} data\footnote{Interactions of gamma photons with the detectors can be of two types: \textit{single events} (SE), when the incoming photon deposits energy in one detector only, and \textit{multiple events} (ME), when the incoming photon deposits energy in several adjacent detectors through Compton diffusion and pair-creation.} with energies in the 506-516\,keV range have been selected and binned into 0.5\,keV-wide bins.

\subsection{Method}
\label{data_method}

\indent Due to the complexity of the response function, the SPI data cannot be inverted and the extraction of astrophysical signals is done by model-fitting. This implies a model for the instrumental background noise and a model for the sky intensity distributions, each model being composed of several component in the general case.\\
\indent In almost all gamma-ray line analyses performed from SPI data so far, the modelling of the background noise is achieved through combinations of \textit{activity tracers} that are hypothesised to reproduce the various trends and time evolutions of the background \citep[for more details, see][]{Jean:2003,Knodlseder:2005,Martin:2009,Martin:2009a}. These tracers are strongly correlated with the data, but since we are looking for signals that are less than 1\% of the data, they need to be finely adjusted through fitting. This approach has proven successful for the study of the nuclear decay lines from $^{26}$Al, $^{60}$Fe and $^{44}$Ti \citep{Wang:2007,Martin:2009,Martin:2009a}, as well as for the study of the diffuse 511\,keV galactic emission \citep{Knodlseder:2005,Weidenspointner:2008}. In the present work, we took advantage of these past achievements and we therefore used the same background modelling as that employed by \citet{Knodlseder:2005} and \citet{Weidenspointner:2008}.\\
\indent The sky models are assumptions on the intensity distribution of the signal under investigation. They can be any kind of observed distribution that the searched emission is expected to be correlated with, or simple analytical models that are astrophysically meaningful and compatible with the performances of the instrument (on angular resolution especially). Each of these intensity distributions is mapped into the data-space through a convolution with the instrument response function (IRF) for each pointing of the data set and then forms a sky model component.\\
\indent The sky model components and the background models are fitted simultaneously to the data using a Maximum Likelihood criterion for Poissonian statistics. The quality of the analysis is then assessed through examination of the residuals. Since we are working with background-dominated data, the modelling of the instrumental background can easily introduce systematic errors that might be compensated for by artificial signals from the sky. In addition, the assumption made on the sky intensity distribution might well be inappropriate or incomplete, which would introduce bias or leave unaccounted events in the data-space.\\
\indent To ensure that our models are adequate, we study the distribution of the residuals  after back-projection on the sky. Basically, for all pointings of the data set, the residuals in each detector are uniformly projected across the mask (or more precisely "across" the instrument response function) over the sky region intersected by the field of view. The distribution of these sky residuals should then follow a statistical distribution, the shape of which is derived from simulated observations. This method is described in greater details in \citet{Martin:2009} and proved for the present case that our hypotheses regarding both the background noise and the sky intensity distribution are satisfactory and that the results obtained are not affected by systematic errors.
\begin{table}[t]
\caption{Positron escape and survival fractions (in \%) as a function of parent nucleus, ejecta mass and transport scheme.}
\begin{center}
\begin{tabular}{|c|c|c|c|c|}
\hline
\celltspace Ejecta mass & Transport & $^{56}$Ni & $^{44}$Ti \cellbspace \\
\hline
\celltspace 2 & Free & 0.4 & 97.4 \cellbspace \\
\hline
\celltspace 2 & Trapped & $< 10^{-2}$ & 90.8 \cellbspace \\
\hline
\celltspace 4 & Free & $< 10^{-2}$ & 94.8 \cellbspace \\
\hline
\celltspace 4 & Trapped & $< 10^{-2}$ & 79.7 \cellbspace \\
\hline
\celltspace 6 & Free & $< 10^{-2}$ & 92.3 \cellbspace \\
\hline
\celltspace 6 & Trapped & $< 10^{-2}$ & 69.0 \cellbspace \\
\hline
\celltspace 8 & Free & $< 10^{-2}$ & 90.0 \cellbspace \\
\hline
\celltspace 8 & Trapped & $< 10^{-2}$ & 59.2 \cellbspace \\
\hline
\end{tabular}
\end{center}
\label{tab_frac}
\end{table}

\subsection{Results}
\label{data_results}

\indent In order to properly assess the 511\,keV emission from the above-listed SNRs, we need to take into account the galactic diffuse emission at this energy to ensure that part of this diffuse flux is not attributed to the SNRs. This is particularly important for G1.9+0.3, which lies very close to the galactic center, where most of the diffuse emission is found. The best-fit model so far for the diffuse annihilation emission is made of three components: a narrow gaussian distribution and a wide gaussian distribution (with typical angular sizes of 3$^{\circ}$ and 11$^{\circ}$ respectively and both centered on the galactic center) and a thick disk \citep{Weidenspointner:2008a}.We thus fit our models for the 6 SNRs simultaneously to the 3-component model for the diffuse emission.\\
\indent Given the $\sim$ 3$^{\circ}$ angular resolution of SPI, all 6 SNRs are mere point-sources for the instrument and we indeed started our analysis with 6 point-sources at the positions of the SNRs plus the 3 components of the diffuse emission. Yet, since positrons could have escaped and propagated away since the explosion, we need to search for annihilation emission not only in the SNRs but also in their surroundings. The largest spatial scale to be considered is $l=ct$, where $t$ is the age of the remnant and $c$ the light-speed. For SN1006 and Tycho, this propagation effect could lead to a substantial increase of the extent of the annihilation emission: up to 15$^{\circ}$ for SN1006 and up to 6$^{\circ}$ for Tycho. We have therefore explored how the result is affected by the source size by replacing the point-source model for these two objects by 2D gaussian profiles of increasing widths.\\
\indent Last, we also explored the potential broadening of the 511\,keV line by analysing the flux in three energy bands of increasing size, starting with the $\sim$ 2\,keV (FWHM) spectral resolution of the instrument at 511\,keV as the minimum size. Significant line broadening could arise from Doppler effect due to the following reasons: 
\begin{enumerate}
\item Annihilation in an expanding medium like the expanding stellar ejecta of a SN/SNR; for a velocity of $\sim$ 1000\,km\,s$^{-1}$, the line broadening is $\sim$ 3\,keV.
\item Annihilation in a hot medium like the shocked regions of SNRs; for a temperature of 10$^{7}$\,K, the broadening is $\sim$ 30\,keV.
\item Annihilation through formation of positronium in flight, which is an important process at the galactic scale and is expected to dominate in the neutral ejecta of SNe; the resultant broadening is $\sim$ 6\,keV \citep{Guessoum:2005,Jean:2006}.
\end{enumerate}
The spectrometric performances of the SPI instrument allow the study of the 511\,keV profile \citep{Jean:2006} but the corollary of such a capability is that too broadened signals become undetectable.\\
\indent The various sky models presented above have been fitted to the allsky SPI data, simultaneously to the background model. No 511\,keV signal was detected from any of the 6 SNRs, whatever the energy band or the spatial extent of the source, in agreement with the latest 511\,keV cartography. The 3-component sky model for the diffuse emission proved robust and was not altered by the addition of the 5 SNRs, especially G1.9+0.3 which could have impacted the output due to its proximity to the galactic center. The added SNR sky models do not lead to any improvement of the fit and thus do not contribute to give a better representation of the SPI data. Finally, we note that using an asymetric model for the galactic diffuse emission \citep[where the disk is split longitudinally in four parts, see][]{Weidenspointner:2008a} gives the same results and does not modify the asymetry observed in the inner Galaxy.\\
\indent From our results, we derived 2$\sigma$ upper-limits on the 511\,keV flux from the 6 SNRs, for the various source sizes and line broadening assumptions. These are listed in Table \ref{tab_obs}. The values reflect the uneven exposure of the sky by INTEGRAL. SN1987A is clearly the least exposed object among the 6 SNRs while G1.9+0.3 is the most exposed. As expected, the limits become less constraining with increasing energy band or source size (because the signal is then diluted in a larger volume of the background-dominated data-space).

\section{Simulations}
\label{simu}

\indent To interpret the above observational constraints, we simulated annihilation lightcurves for a simple description of SN/SNR evolution and a simple uniform CSM/ISM, and under various prescriptions for the transport of the decay positrons. Many analytical formulae developed here were already presented in \citet{Chan:1993}, hereafter CL93, and were just adapted for the purpose of computing annihilation lightcurves.

\subsection{SN/SNR model description}
\label{simu_snr}

\indent In our model, the SN/SNR is represented as a uniform spherical ejecta in homologous expansion. In such a case, the velocity of the stellar material is a linear function of the radius and reaches a maximum $v_{ej}$ at the edge. The evolution of the ejecta is then given by the following relations:
\begin{align}
\label{eq_ejecta1}
& R_{ej} = v_{ej} t = \left( \frac{10}{3} \frac{E_{ej}}{M_{ej}} \right)^{1/2} t \\
\label{eq_ejecta2}
& \rho_{ej} = \frac{3}{4\pi} \frac{M_{ej}}{(v_{ej} t)^{3}} = \frac{3}{4\pi} \left( \frac{3}{10} \right)^{3/2} \frac{M_{ej}^{5/2}}{E_{ej}^{3/2} t^{3}}
\end{align}
where $M_{ej}$ and $E_{ej}$ are the ejecta mass and kinetic energy, which are parameters of the model. $R_{ej}$ and $\rho_{ej}$ are then the ejecta size and density as a function of time $t$.\\
\indent The decay positrons of $^{56}$Ni and $^{44}$Ti are injected at the very centre of the ejecta, where products of explosive Si-burning such as $^{56}$Ni and $^{44}$Ti are expected to be found (yet, strong turbulence during the supernova can alter the ordered chemical stratification that results from a purely spherical explosion). We then consider two extreme transport schemes, as in \citetalias{Chan:1993}: either the positron is trapped in the ejecta, as a result of a strong magnetic turbulence for instance, or it is completely free and can travel radially up to the surface of the ejecta and then escape, provided it is not stopped before.\\
\indent Positrons lose energy along their way through ionisation/excitation of the stellar material assumed to be neutral (but see \ref{simu_adiab}). The energy loss-rate is given by the following law:
\begin{align}
\label{eq_loss}
\frac{dE}{dt} = - \frac{4 \pi r_{0}^{2} m_{e} c^{3} \rho}{u \beta} \left\langle \frac{Z}{A} \right\rangle \left\lbrack \ln \left(\beta \sqrt{\gamma-1} \frac{E+m_{e} c^{2}}{\langle I \rangle} \right) + \phi(E) \right\rbrack
\end{align}
where $E$ is the kinetic energy of the positron, $m_{e} c^{2}$ its mass energy and $\gamma=(1-\beta^{2})^{-1/2}$ its Lorentz factor. The characteristics of the medium are the density $\rho$, the mean ratio of atomic number to mass number $\langle Z/A \rangle$ and the mean ionisation potential $\langle I \rangle$. The constants $r_{0}$ and $u$ are the classical electron radius and the atomic mass unit. The term $\phi(E)$ depends on the particle energy only and its complete expression is given in \citetalias{Chan:1993}. In our model, the medium is assumed to be of a homogeneous chemical composition, with $Z/A=0.5$ and $\langle I \rangle = 100$\,eV, both values being characteristic of the $\alpha$-nuclei that make up most of the ejecta of SNe.\\
\indent If we rewrite Eq. \ref{eq_loss} in a more general way\footnote{The function $\Phi(E)$ actually has a dependence on the chemical composition of the medium (through Z/A and I) but since we work with a homogeneous chemical composition, we dropped that dependence.}:
\begin{align}
\label{eq_slowingdown1}
\frac{dE}{dt} = - \rho \Phi(E) \quad \iff \quad \rho dt = - \frac{dE}{ \Phi(E)}
\end{align}
then the initial and final states of a slowing-down phase, respectively $(t_i,E_i)$ and $(t_f,E_f)$, can be related through:
\begin{align}
\label{eq_slowingdown2}
\int_{t_i}^{t_f} \rho dt = - \int_{E_i}^{E_f} \frac{dE}{ \Phi(E)}
\end{align}
\begin{figure}[!t]
\begin{center}
\includegraphics[width=\columnwidth]{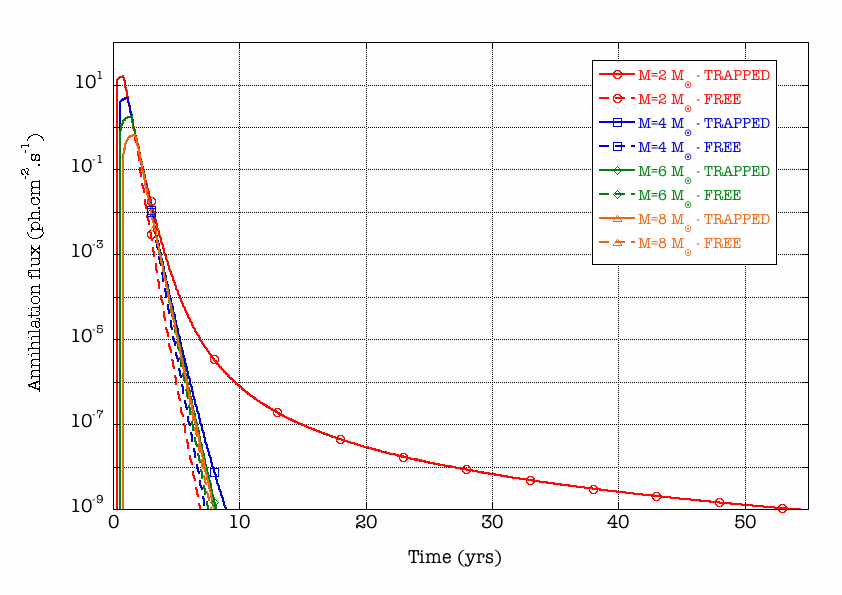}
\includegraphics[width=\columnwidth]{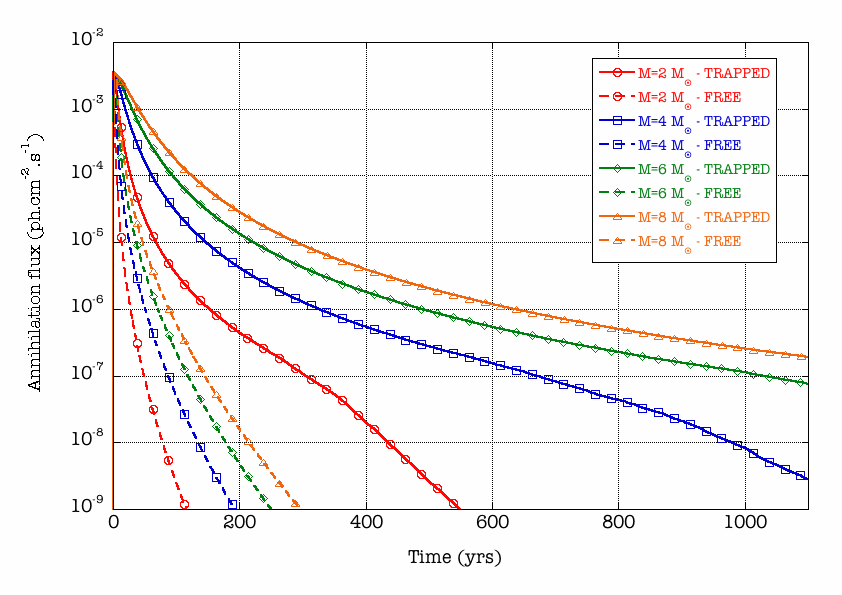}
\caption{511\,keV lightcurve of a SN/SNR for $^{56}$Ni and $^{44}$Ti positrons (upper and lower panel respectively) annihilating in the ejecta, for various ejecta masses and the two transport scenarios.}
\label{lightcurve_SNR}
\end{center}
\end{figure}
The integrated density history experienced by the particle between $t_i$ and $t_f$ is thus related to an integral in the energy space between $E_i$ and $E_f$. Combining Eq. \ref{eq_slowingdown2} with Eq. \ref{eq_ejecta2}, it is possible to compute the time necessary for a freshly injected particle to be slowed down to a certain threshold energy $E_{thr}$, taken in the following to be 100\,eV:
\begin{align}
\label{eq_tsd}
\frac{K}{2} \left(\frac{1}{t_{i}^{2}} - \frac{1}{t_{sd}^{2}} \right)= \int_{E_{thr}}^{E_i} \frac{dE}{ \Phi(E)} \quad \textrm{where} \quad \rho_{ej} = K t^{-3}
\end{align}
In the continuously decreasing density of the expanding ejecta, there is for each injection time a maximum energy above which the particle cannot be slowed down to $E_{thr}$. This energy limit $E_{lim}$ decreases as time passes and is obtained by taking $t_{sd} \to \infty$ in the above equation:
\begin{align}
\label{eq_elim}
\frac{K}{2} \frac{1}{t_{i}^{2}} = \int_{E_{thr}}^{E_{lim}} \frac{dE}{ \Phi(E)}
\end{align}
Practically, this means that an increasing fraction of the positrons released by the decay of $^{56}$Ni or $^{44}$Ti will not be slowed down enough and hence will not annihilate in our model. This is due to our assumption of homologous expansion for the SN/SNR evolution, which allows the density to go to zero at very late times. In reality, however, homologous expansion is halted by the inward propagation of a reverse-shock that basically prevents the ejecta density from falling far below that of the surrounding CSM/ISM. In addition, positrons may escape the SNR and enter this CSM/ISM where their slowing-down proceeds within a finite timescale.\\
\indent In the context of decay positrons in SNe/SNRs, the initial state $(t_i,E_i)$ of a positron is described statistically: the injection time follows an exponential distribution characterised by the half-life of the parent nucleus, while the initial energy obeys a $\beta$-spectrum that depends on the charge of the parent nucleus and on the endpoint energy of the positron \citepalias[the mathematical form of the $\beta$-spectrum is given in][]{Chan:1993}. For each initial state, we compute the slowing-down time $t_{sd}$ using Eq. \ref{eq_tsd}, together with an estimate\footnote{The formula used to estimate $t_{cross}$ assumes that the particle velocity remains constant at its initial value. While moving through the ejecta, the particle velocity decreases so the true crossing time is larger than what we compute. Yet, as shown in \ref{simu_results}, most escaped particles are still relativistic when they exit the ejecta so the error on the crossing time can be considered as small.} of the crossing time $t_{cross}=vt_{i}/(v-v_{ej})$ at which the particle reaches the surface of the ejecta (with $v$ being the particle velocity). Another relevant time is $t_{simu}$, which is the range over which the annihilation lightcurves are computed and is set here at 1100\,yrs. Depending on the obtained values and the adopted transport scheme:
\begin{enumerate}
\item Trapped positrons: If $t_{sd} \leq t_{simu}$, the particle is slowed down within the typical lifetime of a young SNR and it annihilates at a time $t_{sd}$.
\item Trapped positrons: If $t_{sd} > t_{simu}$, the particle cannot be slowed down and it survives inside the ejecta.
\item Free positrons: If $t_{sd} \leq t_{cross} \leq t_{simu}$, the particle is slowed down before it can escape the ejecta and it annihilates at a time $t_{sd}$.
\item Free positrons: If $t_{sd} > t_{cross}$, the particle can escape into the CSM/ISM the ejecta before it is completely slowed down.
\end{enumerate}
In the above scenario, we considered that once slowed down to the threshold energy, the positrons all annihilate through positronium formation in-flight in the ejecta assumed to be mostly neutral (see \ref{decaypositrons_sne}). The process of positronium formation and subsequent annihilation is considered to be immediate owing to its very large cross-section and the ns-$\mu$s lifetime of the bound-state.\\
\indent The model described above yields the time distributions of annihilation for $^{56}$Ni or $^{44}$Ti positrons, for a given ejecta mass and energy and for the two transport schemes. SN/SNR lightcurves at 511\,keV are eventually obtained after normalisation by the following factor:
\begin{equation}
\label{eq_norm}
f_{norm} = \frac{Y_{A} f_{\beta}}{Au} \frac{2 f_{Ps}}{4} \frac{1}{4 \pi d^{2}} \frac{1}{dt} 
\end{equation}
where $Y_{A}$ and $A$ are respectively the typical yield of the radionuclei and its mass number, and $f_{\beta}$ is the branching fraction of $\beta^+$-decay for that isotope. Then, $f_{Ps}$ is the assumed positronium fraction and the factors 1/4 and 2 account for the fact that 25\% of the positroniums are parapositroniums, which decay by emitting 2 photons at 511\,keV. Last, $d$ is the distance to the SNR and $dt$ is the size of time bins in our simulation.\\
\indent The transport of 511\,keV photons in the SN/SNR is taken into account in a very basic way, by comparing the mean free path of the photons in the uniform density ejecta to the radius of the ejecta. In the computation of the 511\,keV photon mean free path, we used a value of 8.63\,10$^{-2}$\,cm$^{2}$\,g$^{-1}$ for the attenuation factor, which formally corresponds to the O element but also applies to elements from Ne to Ni\footnote{The attenuation factor values were taken from the Xcom database of the National Institute of Standards and Technology; see "http://physics.nist.gov/PhysRefData/Xcom/html/xcom1.html"}.\\
\indent In the process, we also computed the spectra of the escaped particles as these will be the input to the modelling of the annihilation in the surrounding CSM/ISM. The energy $E_{esca}$ of an escaped particle is given by:
\begin{align}
\label{eq_nrjesca}
\frac{K}{2} \left(\frac{1}{t_{i}^{2}} - \frac{1}{t_{cross}^{2}} \right)= \int_{E_{esca}}^{E_i} \frac{dE}{ \Phi(E)}
\end{align}
As we will see later, escaped particles in our model have a mean kinetic energy that is a few 100\,keV below the mean kinetic energy at decay, which contributes to an enhanced annihilation in the surrounding CSM/ISM.

\subsection{CSM/ISM model description}
\label{simu_ism}

\begin{figure}[t]
\begin{center}
\includegraphics[width=\columnwidth]{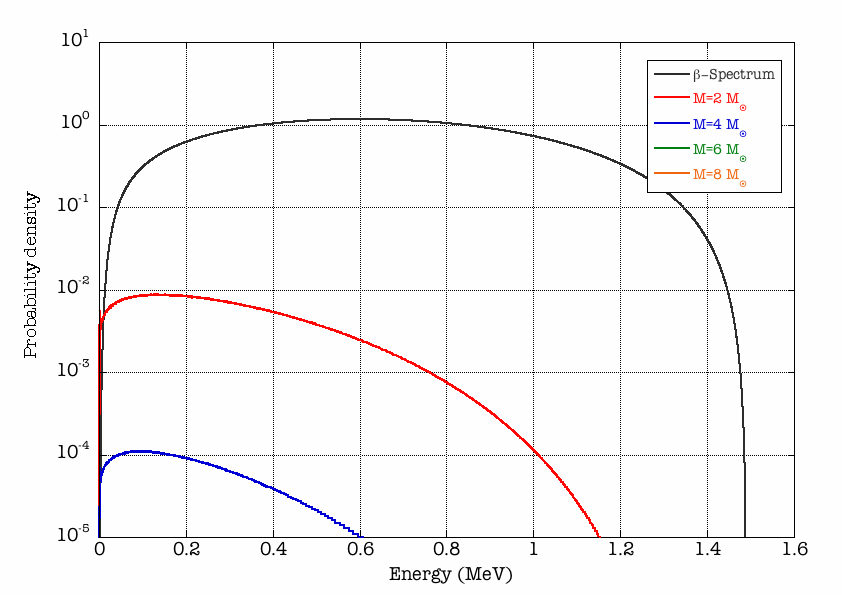}
\includegraphics[width=\columnwidth]{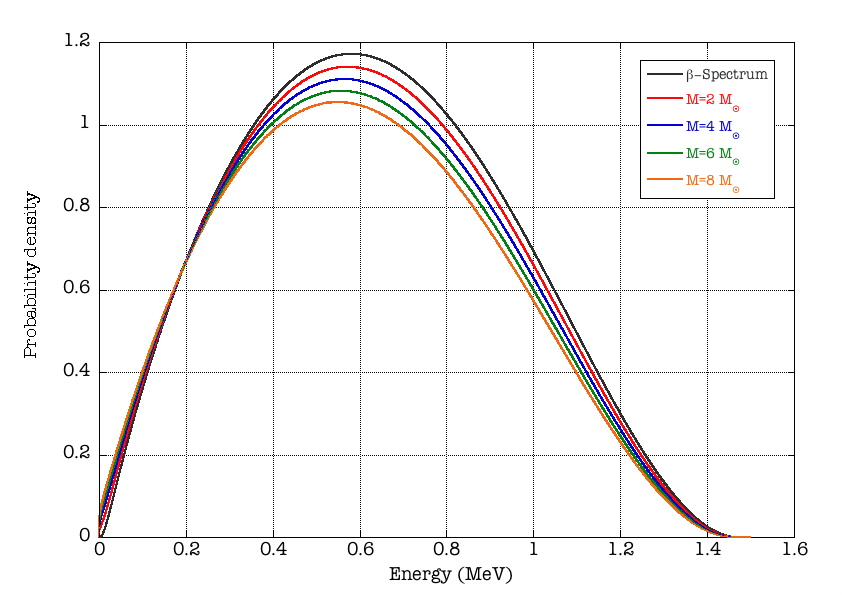}
\caption{Spectra of the escaping $^{56}$Ni and $^{44}$Ti positrons (upper and lower panel respectively), as a function of ejecta mass. Also shown for comparison is the original $\beta$-spectrum.}
\label{spec_SNR}
\end{center}
\end{figure}
The model for the CSM/ISM is actually a simplified version of that used for SNe/SNRs. The density in the medium surrounding the SN/SNR is assumed to be constant, which results in a more simple equation to compute the slowing-down time:
\begin{align}
\label{eq_tsd_ism}
\rho_{0}(t_{sd}-t_{i})= \int_{E_{thr}}^{E_i} \frac{dE}{ \Phi(E)}
\end{align}
Then, the methodology to compute annihilation lightcurves is the same as for the case of SNe/SNRs, except that an escape fraction is added to the normalisation factor of Eq. \ref{eq_norm}. Another difference is that the energy distribution of the positrons injected in the CSM/ISM may differ from the $\beta$-spectrum that prevailed inside the SN/SNR because the particles already lost a fraction of their initial energy on their way out.\\
\indent If the CSM/ISM is assumed to be neutral, energy loss proceeds through ionisation/excitation and annihilation again occurs mainly by positronium formation in-flight immediately after slowing-down. If the CSM/ISM is assumed to be strongly ionised, energy loss proceed through Coulomb losses, which are $\sim$10 times more efficient than excitation/ionisation and would therefore favour earlier annihilation. Yet, if the CSM/ISM is assumed to be fully ionised, no H atoms exist for positronium formation in-flight and annihilation occurs mostly through positronium formation by radiative recombination of thermalised positrons with free electrons. This process, however, has quite a small cross-section and takes place over characteristic timescales of $\sim 10^{5}$ $n_e^{-1}$\,yrs (where $n_e$ is the electron density), which strongly limits the annihilation in the vicinity of the SNR and rather favours diffusion out to large distances. In a warm ionised medium, cosmic interstellar He is essentially neutral \citep{Ferriere:2001}, but positronium formation by charge exchange with He is negligible as positrons are preferentially slowed down below the energy threshold of the process by the free electrons. Similarly, annihilation in dust grains and on PAHs contributes only marginally mainly because of their very low densities \citep{Guessoum:2006}. On the whole, it seems that a strongly ionised medium is not a favourable condition for annihilation on timescales of the order of the typical ages of young SNRs, and we will therefore consider in the following only slowing-down and annihilation in a predominantly neutral CSM/ISM.
\begin{figure}[t]
\begin{center}
\includegraphics[width=\columnwidth]{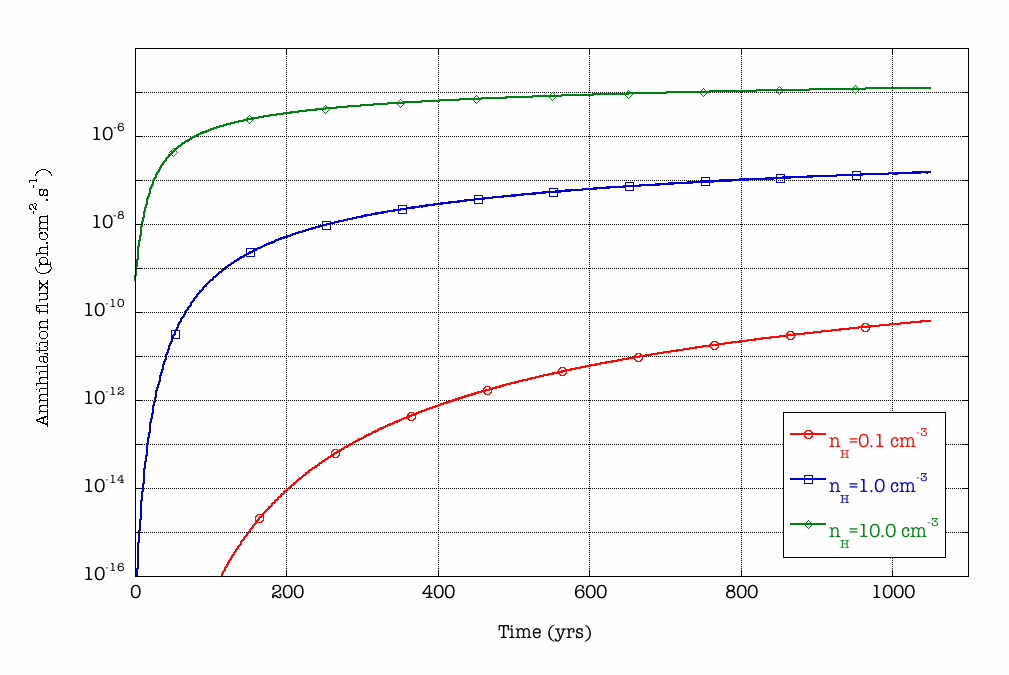}
\includegraphics[width=\columnwidth]{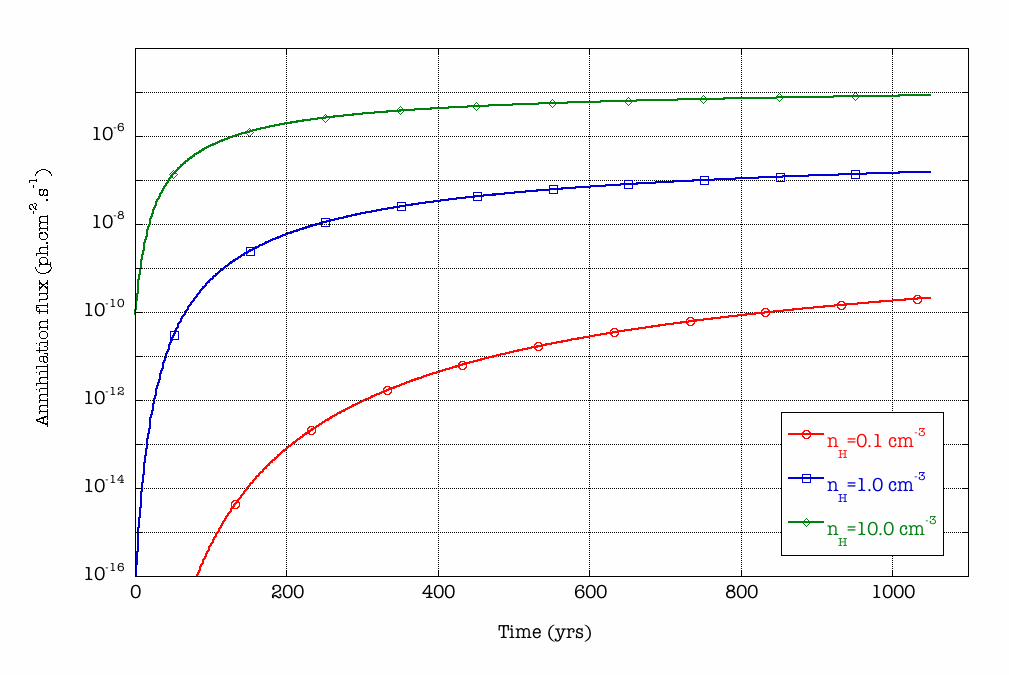}
\caption{511\,keV lightcurve of a SN/SNR for $^{56}$Ni and $^{44}$Ti positrons (upper and lower panel respectively) annihilating in the surrounding CSM/ISM after escape, for various atomic hydrogen densities. The energy distribution of the particles is their original $\beta$-spectrum.}
\label{lightcurve_ISM}
\end{center}
\end{figure}

\subsection{Results}
\label{simu_results}

\indent Based on the above model, we computed SN/SNR lightcurves at 511\,keV for both $^{56}$Ni and $^{44}$Ti positrons, for a single canonical explosion energy of 10$^{51}$\,erg and a range of ejecta masses, and under the two extreme prescriptions for positron transport. In order to allow an easy scaling of our results (so as to apply it to specific cases), we used isotope yields of 10$^{-1}$\,M$_{\odot}$ and 10$^{-4}$\,M$_{\odot}$ for $^{56}$Ni and $^{44}$Ti respectively, together with an assumed positronium fraction of 90\% and a distance to the SN/SNR of 1\,kpc. The results are presented in Fig. \ref{lightcurve_SNR}.\\
\indent Whatever the transport scenario, the annihilation lightcurves for $^{56}$Ni positrons quickly rise (within a year) to quite high fluxes, with a peak value of $\sim$1-10\,ph\,cm$^{-2}$\,s$^{-1}$ (for a distance of 1\,kpc), that is about 10$^5$-10$^6$ times the present-day sensitivity of INTEGRAL/SPI for point-sources (see Table \ref{tab_obs}). After that "annihilation flash", the emission rapidly drops. Over the first years, the lightcurves closely follow the decay of $^{56}$Ni, which can be explained by a short slowing-down time with respect to the characteristic time of the decay. Then, the lightcurves depart from the decay curve as the lifetime of the particles becomes long enough for an increasing fraction of the free positrons to escape, while the trapped positrons annihilate with an increasing delay\footnote{In the trapped case, one may have the impression that the lightcurves are non-physical since after following the decay curve, they rise above it thereby suggesting that more positrons are annihilated than created. This is a misleading effect of the many orders of magnitude spanned by the lightcurves. Actually, only a tiny fraction of the positrons created over the first years annihilate with a delay and the delayed emission a few decades after the explosion is accordingly tiny compared to the flux levels of the early years.}. Lighter ejecta favour an early escape and the corresponding annihilation lightcurves for free positrons therefore drop more rapidly. In the case of trapped positrons, the lower densities associated with the very lightest ejecta imply longer slowing-down times and hence a higher flux at later times.\\
\begin{figure}[t]
\begin{center}
\includegraphics[width=\columnwidth]{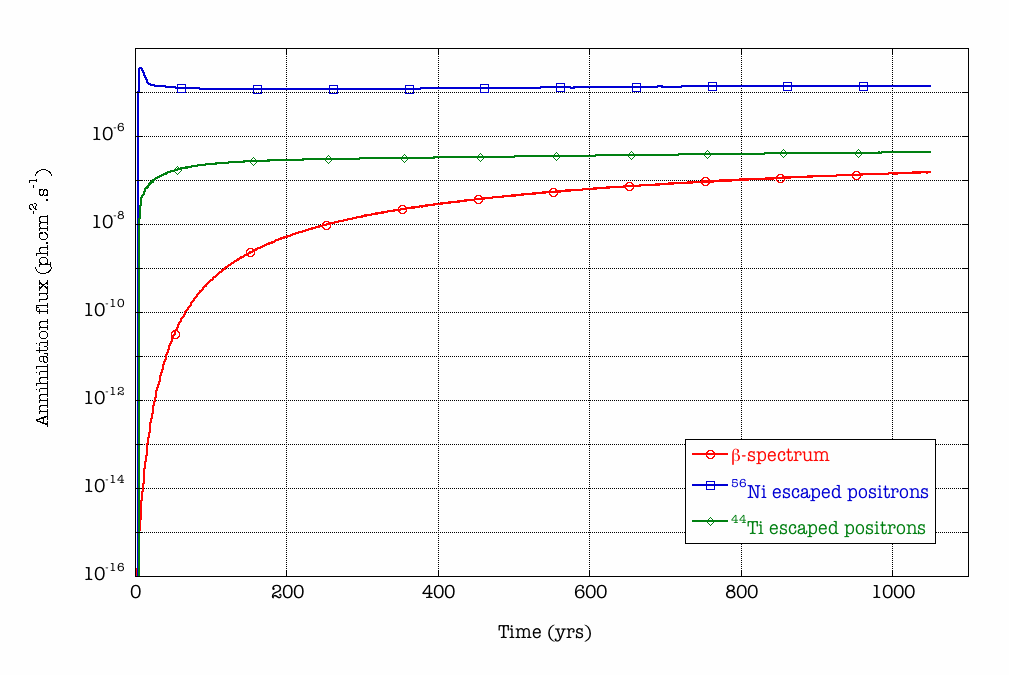}
\caption{511\,keV lightcurve of the CSM/ISM for $^{56}$Ni or $^{44}$Ti positrons from SNe/SNRs, for an atomic hydrogen density of 1\,cm$^{-3}$ and depending on the energy distribution of the particles.}
\label{lightcurve_ISM_esca}
\end{center}
\end{figure}
\indent The annihilation lightcurves for $^{44}$Ti seem to follow similar trends. Nevertheless, the continous injection of $^{44}$Ti positrons over centuries, as opposed to the burst of $^{56}$Ni positrons, leads to some differences. In particular, the delayed emission of trapped positrons increases with ejecta mass, contrary to the case of $^{56}$Ni positrons. In the latter situation, almost all positrons are released at early times and need to survive as long as possible for substantial late emission to exist, which requires light and thin ejecta; in contrast, most $^{44}$Ti positrons are released at intermediate times, when the ejecta is already considerably diluted, and the key for late emission is then an efficient slowing-down, which requires thick and massive ejecta. On the whole, the most favourable conditions to observe 511\,keV emission from young Galactic SNRs therefore seem to be a massive ejecta, a high yield of $^{44}$Ti and enough magnetic turbulence so as to confine the decay positrons inside the ejecta.\\
\indent The escape fractions obtained in the case of free positrons are listed in Table \ref{tab_frac}, together with the survival fractions obtained in the case of trapped positrons. The $\sim 0$\% escape fractions for $^{56}$Ni positrons agree with the negligible values found by \citetalias{Chan:1993} for similar ejecta masses and no mixing (their models B4B, B6C and B8A can be approximately compared to our 2, 4 and 6\,M$_{\odot}$ ejecta models). \citetalias{Chan:1993} showed that higher escape fractions of a few \% can be obtained if the $^{56}$Ni is mixed out in the ejecta, for higher explosion energies, or in the case of lighter ejecta masses (of SNIa for instance), but we did not explore these effects here.\\
\indent The spectra of the escaped free particles are shown in Fig. \ref{spec_SNR} for the two parent radio-isotopes (in the case of $^{56}$Ni, the escape fractions are so small that the spectra are shown on a logarithmic scale, for the 2 and 4\,M$_{\odot}$ ejecta only). The original $\beta$-spectrum is also shown for comparison. In both cases, the number of escaped particles decreases as ejecta mass increases and their spectrum is shifted to lower and lower energies. This will have appreciable consequences for the annihilation in the CSM/ISM.\\
\indent The 511\,keV lightcurves for the annihilation of decay positrons in the CSM/ISM surrounding the SNR are shown in Fig. \ref{lightcurve_ISM}, for various atomic hydrogen densities. Again, in order to allow an easy scaling of our results, we used isotope yields of 10$^{-1}$\,M$_{\odot}$ and 10$^{-4}$\,M$_{\odot}$ for $^{56}$Ni and $^{44}$Ti respectively, together with an assumed positronium fraction of 90\% and a distance to the SN/SNR of 1\,kpc. In addition, the escape fractions are assumed to be 1\% for $^{56}$Ni positrons and 100\% for $^{44}$Ti positrons.\\
\indent In Fig. \ref{lightcurve_ISM}, the positrons were injected with an energy distribution corresponding to their original and unaltered $\beta$-spectrum and the lightcurves obtained for $^{56}$Ni and $^{44}$Ti positrons are remarkably similar. The fact that nearly identical levels of emission are reached can actually be explained by the mean characteristics of each positron source: although the yield of $^{56}$Ni is on average $10^{3}$ higher than the yield of $^{44}$Ti, only 20\% of the $^{56}$Ni nuclei undergo a $\beta^+$-decay and because of the short lifetime of the isotope only $\sim$ 1\% of the positrons can escape; in contrast, almost all $^{44}$Ti nuclei produce positrons and almost all these positrons escape.\\
\indent It should also be noted that the evolution of the annihilation lightcurves with density agrees with expectations. If a flux $f_{511}$ is obtained at $t$ for a CSM/ISM density $n_1$, a flux $(n_2/n_1)f_{511}$ is obtained at $(n_1/n_2)t$ for a CSM/ISM density $n_2$ (because the slowing-down time decreases as $n$ increases, and so successive positron annihilations in a higher density medium occurs sooner and in a smaller time interval). The scaling works as expected for $^{56}$Ni, but not so well for $^{44}$Ti due to its longer decay time (which prevents, for instance, the comparison of the flux at 1000\,yrs in the $n=0.1$\,cm$^{-3}$ case with the flux at 100\,yrs in the $n=1.0$\,cm$^{-3}$ case because all positrons have been released at 1000\,yrs but not at 100\,yrs).\\
\indent Prior to entering the CSM/ISM, however, most positrons lost a fraction of their initial energy on their way through the ejecta (see Fig. \ref{spec_SNR}). When these modified spectra are taken into account, the resulting annihilation in the CSM/ISM is noticeably different. In Fig. \ref{lightcurve_ISM_esca}, the lightcurves obtained for $^{56}$Ni and $^{44}$Ti positrons entering a 1\,cm$^{-3}$ CSM/ISM after crossing a 4\,M$_{\odot}$ ejecta are shown, together with the unaltered $\beta$-spectrum case (a single curve is shown for both isotopes since, as discussed earlier, they are quite similar). It appears that the annihilation flux from the CSM/ISM can be increased by up to two orders of magnitude in the case of $^{56}$Ni positrons when the slowing-down of the particles inside a 4\,M$_{\odot}$ ejecta is taken into account. We want to emphasise, however, some simplifications made when computing the CSM/ISM lightcurves with modified positron spectra. First, we used escape fractions of 1\% and 100\% for $^{56}$Ni and $^{44}$Ti positrons respectively, but these are not the values given by our model for a 4\,M$_{\odot}$ ejecta (see Table \ref{tab_frac}). We did that to ease the comparison with Fig. \ref{lightcurve_ISM}, but one should keep in mind that the escape fraction and the spectrum of escaped particles are linked (higher escape fractions are associated with more energetic escaping particles). Then, we used the same modified spectrum for all positrons injected in the CSM/ISM, but this spectrum actually evolves with time (the mean energy of escaping particles increases with time, as the ejecta becomes more and more transparent). Nevertheless, the comparison of simulations made for different modified spectra and different injection time profiles showed that these simplifications only have a modest impact on the predicted lightcurves.
\begin{table*}[!t]
\begin{minipage}[][8cm][c]{\textwidth}
\renewcommand{\footnoterule}{}
\renewcommand{\thefootnote}{\alph{footnote}}
\begin{center}
\caption{Characteristics of the six SNRs used to compute their 511\,keV annihilation lightcurves. Numbers in italic are typical values assumed when no established measurements were found in the literature.}
\begin{tabular}{|c|c|c|c|c|c|c|}
\hline
\celltspace  & Cas A & Tycho & Kepler & SN1006 & SN1987A & G1.9+0.3 \cellbspace \\
\hline
\celltspace Age (yrs) & 338 & 437 & 405 & 1003 & 22 & 100 \footnotemark[1] \cellbspace \\
\hline
\celltspace Distance (kpc) & 3.4 \footnotemark[2] & 2.4 \footnotemark[3] & 6 \footnotemark[4] & 2.2 \footnotemark[5] & 50 & 8.5 \footnotemark[1] \cellbspace \\
\hline
\celltspace Type & SNIIb \footnotemark[6] & SNIa & SNIa & SNIa & SNII & SNIa \footnotemark[1] \cellbspace \\
\hline
\celltspace Ejecta mass (M$_{\odot}$) & 2.2 \footnotemark[7] & \textit{1.4} & \textit{1.4} & \textit{1.4} & 14 \footnotemark[8] & \textit{1.4} \cellbspace \\
\hline
\celltspace Explosion energy (10$^{51}$\,erg) & 2.0-4.0 \footnotemark[9] & \textit{1} & \textit{1} & \textit{1} & 1.1 $\pm$0.3 \footnotemark[8] & \textit{1} \cellbspace \\
\hline
\celltspace $^{56}$Ni yield (M$_{\odot}$) & 0.07-0.15  \footnotemark[6] & \textit{0.6} & \textit{0.6} & \textit{0.6} & 0.07 & \textit{0.6} \cellbspace \\
\hline
\celltspace $^{44}$Ti yield (M$_{\odot}$) & 1.6\,10$^{-4}$ \footnotemark[10] &  $< 2\,10^{-4}$ \footnotemark[11] & $< 3\,10^{-5}$ \footnotemark[12] & $< 3\,10^{-5}$  \footnotemark[12] & 2\,10$^{-4}$ & $< 3\,10^{-5}$ \footnotemark[12] \cellbspace \\
\hline
\celltspace CSM density (cm$^{-3}$) & 3.2 \footnotemark[13] & 0.3 \footnotemark[13] & 0.5 \footnotemark[4]  & 0.1 \footnotemark[14] & 0.02 \footnotemark[15] & 0.04 \footnotemark[1]  \cellbspace \\
\hline
\end{tabular}
\label{tab_SNRs}
\end{center}
References: (a) \citet{Reynolds:2008} (b) \citet{Reed:1995} (c) \citet{Cassam-Chenai:2007} (d) \citet{Vink:2008} (e) \citet{Winkler:2003} (f) \citet{Krause:2008} (g) \citet{Willingale:2003} (h) \citet{Blinnikov:2000} (i) \citet{Laming:2006} (j) \citet{Renaud:2006} (k) \citet{Renaud:2009} (l) \citet{Renaud:2006a} (m) \citet{Truelove:1999} (n) \citet{Raymond:2007} (o) \citet{Sugerman:2005}
\end{minipage}
\end{table*}

\subsection{Adiabatic losses}
\label{simu_adiab}

\begin{figure}[t]
\begin{center}
\includegraphics[width=\columnwidth]{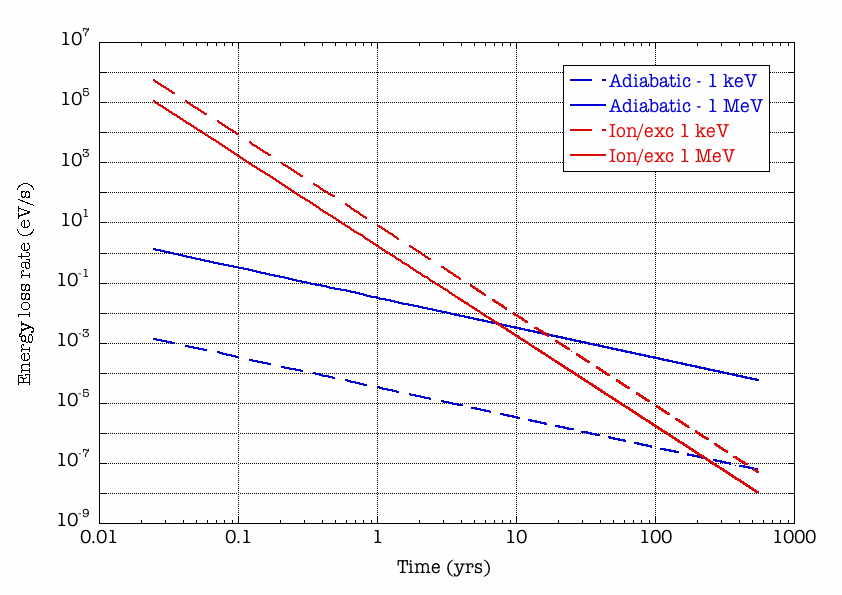}
\caption{Adiabatic losses and ionisation/excitation losses as a function of time for a 2\,M$_{\odot}$ ejecta and for 1\,keV and 1\,MeV particle energies.}
\label{fig_adiab}
\end{center}
\end{figure}

\indent In our model of positron transport in expanding SN/SNR ejecta, we assumed that the particles lose energy through ionisation/excitation of the stellar material. Theoretically, however, positrons also suffer from adiabatic losses. In the following, we discuss how these additional losses may impact the predicted lightcurves.\\
\indent The general expression for adiabatic losses is given by Eq. \ref{eq_adiab_gen}, where $E$ is the energy of the gas particles, $\gamma$ the ratio of specific heats and $\vec{v}$ the velocity of the flow.
\begin{equation}
\label{eq_adiab_gen}
\frac{dE}{dt}= -(\gamma -1) \left( \vec{\nabla} . \vec{v} \right)\,E
\end{equation}
In the specific case of homologous expansion ($v=r/t$) of an ultrarelativistic gas ($\gamma=4/3$), this equation becomes:
\begin{equation}
\label{eq_adiab_homo}
\frac{dE}{dt}= -\frac{E}{t} \\
\end{equation}
In contrast to energy losses by ionisation/excitation, energy losses by adiabatic expansion do not depend on the characteristics of the ejecta (such as mass or kinetic energy) and they increase with particle energy. In order to be efficiently slowed down by adiabatic losses, however, the particles need to be confined to the expanding volume for a sufficiently long time compared to the expansion timescale.\\
\indent In Fig. \ref{fig_adiab} are plotted the energy losses by both processes as a function of time, for an ejecta mass of 2\,M$_{\odot}$ and for particle energies of 1\,keV and 1\,MeV. From this comparison, it appears that adiabatic losses can be neglected over the first years after explosion. Then, owing to the fact that ionisation/excitation losses decrease as $1/t^3$ while adiabatic losses decrease as $1/t$ only, the latter become progressively dominant. Energy losses by adiabatic expansion take over after $\sim$5-10\,yrs for $\sim$MeV particles and after $\sim$500-1000\,yrs for $\sim$keV particles. For higher ejecta masses, the shift occurs later as the associated higher densities (for a same explosion energy) maintain ionisation/excitation losses over adiabatic losses for a longer time (ionisation/excitation losses scale with density, which scales as the power 5/2 of ejecta mass; see Eq. \ref{eq_ejecta2}).\\
\indent In the context of our simple model, it is not possible to include both processes in Eq. \ref{eq_slowingdown2} to compute the slowing-down time because of their different time dependence. Nevertheless, we show below from simple arguments that adiabatic losses are not expected to modify the annihilation lightcurves computed from our model.\\
\indent In \ref{simu_snr}, we showed that there exists for each particle injected in the ejecta at time $t$ an energy limit $E_{lim}(t)$ above which the particle cannot be slowed down in a finite time (see Eq. \ref{eq_elim}). Under the assumption of an undisturbed ejecta expansion, this energy limit gives the increasing fraction of trapped positrons that will survive in the ejecta, mostly as relativistic particles. It also gives to a very good approximation the increasing fraction of free positrons that will escape the ejecta. Although not really intuitive, it can be easily shown that for each time $t$ after explosion, the particle energy above which adiabatic losses dominate over ionisation/excitation losses has almost the same expression as $E_{lim}$ (one just need to equate both energy loss formulae to get the result). In a very opportune way, adiabatic losses become important at a given energy when ionisation/excitation losses fail to efficiently slow down positrons with that initial energy (and this works whatever the ejecta mass and explosion energy). At this point, the energy loss rate of adiabatic expansion and ionisation/excitation are almost equal, but because ionisation/excitation losses decline faster than adiabatic losses, the particle rapidly suffers from adiabatic losses only. From integration of Eq. \ref{eq_adiab_homo}, the slowing-down time for a particle injected at time $t_i$ with initial energy $E_i$ and suffering adiabatic losses is found to be\footnote{Strictly speaking, Eq. \ref{eq_adiab_evol} should include a numerical factor of the order of a few because the energy loss rate of Eq. \ref{eq_adiab_homo} increases by up to a factor of 2 as particles turn from ultra-relativistic to mildly or non-relativistic.} :
\begin{equation}
\label{eq_adiab_evol}
t_{sd}= \frac{E_i}{E_{thr}}\,t_i \\
\end{equation}
From Eq. \ref{eq_adiab_evol} and Fig. \ref{fig_adiab}, we can now estimate how positrons suffering adiabatic losses can affect the annihilation lightcurves for the case of a 2\,M$_{\odot}$ ejecta. Positrons with initial energy of $\sim$1\,MeV (respectively $\sim$1\,keV) are affected by adiabatic losses from 5-6\,yrs (respectively 500-600\,yrs) after explosion, and they are slowed down to $E_{thr}$ after 50000-60000\,yrs (respectively 5000-6000\,yrs). Adiabatic losses in a 2\,M$_{\odot}$ ejecta expanding with a kinetic energy of 10$^{51}$\,erg therefore lead to annihilation at very late times, well beyond the typical ages of the SNRs in our sample. Moreover, light ejecta are the most favourable case because adiabatic losses take over more rapidly, and the corresponding annihilation occurs at the soonest. We can thus safely conclude that adiabatic losses would impact our predicted annihilation lightcurves only marginally.

\section{Discussion}
\label{discu}

We compared our upper-limits on the 511\,keV emission from the youngest local SNRs with the expected fluxes computed in the frame of our simple models. In this comparison, we took into account the peculiarities of each object that can impact the annihilation emission.

\subsection{Characteristics of the young SNRs}
\label{discu_SNRs}

\indent The main parameters required to model the annihilation in an individual SNR are the ejecta mass, the $^{56}$Ni and $^{44}$Ti yields, the density of the surrounding medium and the distance to the object. In Table \ref{tab_SNRs} are listed the relevant characteristics of the SNRs considered in this study. The quantities for which we did not find clearly established measurements in the literature were assumed to have typical values.\\
\indent The ejecta mass for SNIa explosions was set to 1.4\,M$_{\odot}$, with the implicit assumption that these events correspond to the deflagration or detonation of a white dwarf that reached the Chandrasekhar mass through accretion of matter from a companion star. For Cas A, the ejecta mass was determined from modelling of the X-ray emission of the remnant, while for SN1987A it was determined from modelling of the optical/UV lightcurve of the supernova over the first months.\\
\indent The $^{56}$Ni yield of all SNe Ia was set to 0.6\,M$_{\odot}$, as deduced from the quite homogeneous early lightcurves of this type of events. The $^{56}$Ni yield of Cas A was assumed to be in the range inferred for SN1993J because the optical spectrum near peak brightness of Cas A, echoed by interstellar dust several centuries after the explosion, is quite similar to the optical spectrum of SN1993J, which is taken to be a canonical SNIIb. In contrast, the $^{56}$Ni yield of SN1987A was deduced directly from its early lightcurve.\\
\indent The production of $^{44}$Ti by SNe remains quite controversial because Cas A is currently the only observed source of gamma-ray radiation from $^{44}$Ti decay, while emission from younger and probably obscured SNRs is statistically expected \citep[see the discussion in][]{The:2006}. The $^{44}$Ti yield of the Cas A explosion was estimated to 1.6\,10$^{-4}$\,M$_{\odot}$ from direct observations of the gamma-ray decay lines. A similar value was indirectly inferred for SN1987A from the modelling of the late lightcurve of the event. For Tycho and G1.9+0.3, upper-limits on the flux at the energy of the two hard X-ray decay lines at 68 and 78\,keV were translated into 3$\sigma$ upper-limits on the mass of $^{44}$Ti synthesised in the explosions. These limits are consistent with most theoretical yields obtained in deflagration and delayed detonation models of SNIa explosions, which range from 0.8 to 4.6\,10$^{-5}$\,M$_{\odot}$ \citep{Iwamoto:1999}. For all SNe Ia, we therefore adopted a yield of 3\,10$^{-5}$\,M$_{\odot}$, in agreement with the predictions and the strongest constraint set by G1.9+0.3.\\
\indent The CSM/ISM densities around SNe/SNRs can be estimated by various methods. From a modelling of its expansion dynamics, Cas A was found to be expanding in the dense wind of its red supergiant progenitor. The same method was applied to Tycho\footnote{We note that, for Tycho, the 0.3\,cm$^{-3}$ density derived by \citet{Truelove:1999} from dynamical arguments is consistent with the 0.6\,cm$^{-3}$ limit obtained by \citet{Cassam-Chenai:2007} to account for the absence of thermal X-ray emission from the shocked ambient gas}, Kepler and G1.9+0.3. For SN1006, the 0.1\,cm$^{-3}$ CSM/ISM density is imposed by the observed thickness of a Balmer-dominated filament of the forward shock. For SN1987A, the density we adopted corresponds to the average value obtained from the $\sim$2\,M$_{\odot}$ of material that were estimated from light-echoes to form the circumstellar environment of the progenitor within about 30 light-years.\\
\indent Most of these works, however, rely on the assumption of a constant density environment while the actual density structure around a SNR is quite often inhomogeneous and evolves with distance. In the general case, the density structure around a core-collapse SN results from the interaction of successive stellar outflow episodes, each with a specific mass loss rate and velocity. In the spherical case, this can give rise to a series of concentric shells of various densities and thicknesses \citep{Garcia-Segura:1996,Garcia-Segura:1996a,Dwarkadas:2005}, but reality seems to be even more complicated. Light-echoes from SN1987A have revealed the complex circumstellar environment shaped by the progenitor B-star before its explosion: the slow and dense wind blown during the red supergiant phase is thought to have been asymmetric and, while expanding in the matter expelled during Main-Sequence, created an equatorial overdensity that forced the subsequent blue supergiant wind to polar directions. The resulting double-lobed structure spans several orders of magnitude in density and is the proposed explanation for the three rings that appeared shortly after the supernova \citep[see][and references therein]{Sugerman:2005}. The case of thermonuclear SNe is not less complicated. During the pre-SN evolution, as the white dwarf grows in mass through accretion, the surrounding environment can be shaped by a variety of processes among which the mass-loss from a red giant companion, a fast accretion wind or recurrent nova explosions \citep{Badenes:2007,Hachisu:2008,Borkowski:2009}.\\
\indent In addition to the uncertainties on the density structure around SNe/SNRs, the physical state of the surrounding gas is also relevant to the annihilation of decay positrons. As explained previously, a strongly ionised medium is more efficient at slowing down the escaping positrons but then implies annihilation on very long timescales (compared to the typical ages of young SNRs). The massive star progenitors of core-collapse SNe are powerful sources of UV radiation that ionise the surrounding medium over large distances. In addition, the burst of soft X-rays and extreme UV at supernova shock breakout ionise or reionise the CSM/ISM up to certain distance. The timescale for recombination in the CSM/ISM may be longer than the age of the SNRs and so escaping positrons may not annihilate efficiently over the first centuries.

\subsection{Observations versus predictions}
\label{discu_compa}

\begin{figure}[t]
\begin{center}
\includegraphics[width=\columnwidth]{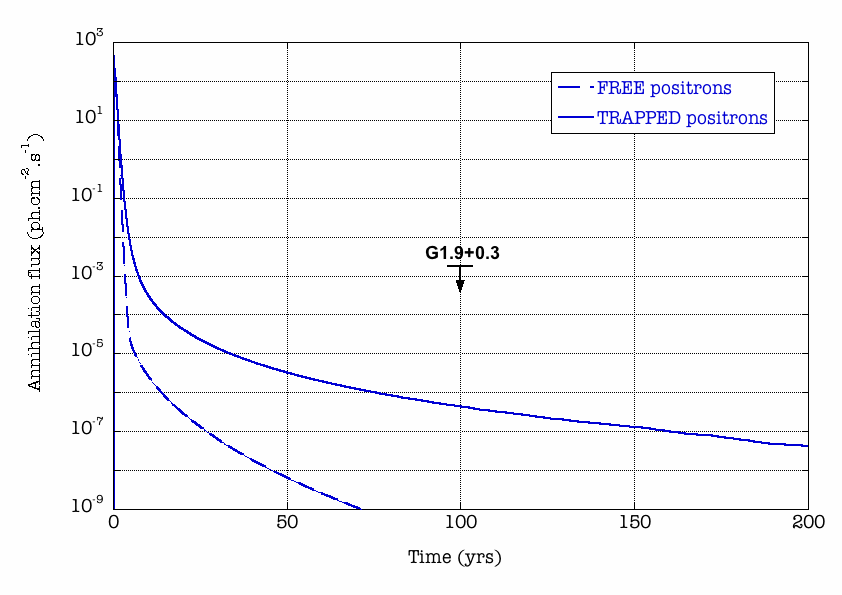}
\caption{Total 511\,keV lightcurve from a SNIa at a 1\,kpc distance, combining the contribution of $^{56}$Ni and $^{44}$Ti positrons. Also shown are the upper-limits on the 511\,keV flux from INTEGRAL/SPI.}
\label{lightcurve_SNIa}
\end{center}
\end{figure}

\begin{figure}[!t]
\begin{center}
\includegraphics[width=\columnwidth]{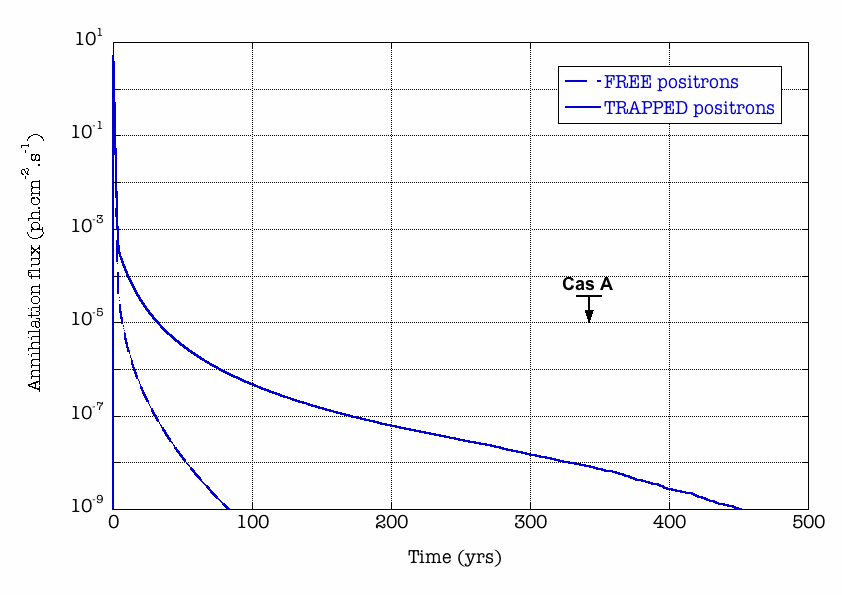}
\includegraphics[width=\columnwidth]{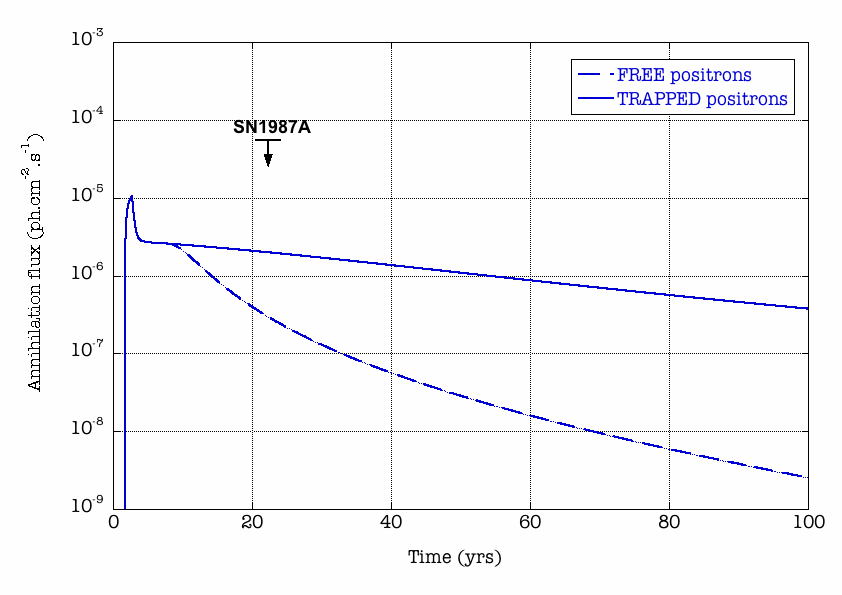}
\caption{Total 511\,keV lightcurve for the two remnants of core-collapse SNe, Cas A and SN1987A, combining the contribution of $^{56}$Ni and $^{44}$Ti positrons. Also shown are the upper-limits on the 511\,keV flux from INTEGRAL/SPI.}
\label{lightcurve_ccSNe}
\end{center}
\end{figure}

\indent From the data listed in Table \ref{tab_SNRs}, we computed the expected fluxes for each SNR. As observational constraints, we considered the INTEGRAL/SPI upper limits on the 511\,keV flux obtained in a 5\,keV band around the line, because annihilation in our model was taken to occur mostly through positronium formation in-flight, which causes a Doppler broadening of about 6\,keV (see \ref{data_results}). Moreover, we assumed the most constraining case of no spatial diffusion. In the following, we discuss separately the case of annihilation in the expanding ejecta and in the CSM/ISM.\\
\indent Instead of showing specific synthetic 511\,keV lightcurves for each remnant of a SNIa, we took advantage of the homogeneity of this class of object to model a single set of lightcurves for a paradigmatic SNIa located at a distance of 1\,kpc. Our upper-limits on the annihilation fluxes were then scaled to that distance and compared to the predictions.\\
\indent The resulting lightcurves, including the contributions from both $^{56}$Ni and $^{44}$Ti positrons and depending on the transport scenario, are shown in Fig. \ref{lightcurve_SNIa} together with the constraints obtained from INTEGRAL/SPI observations. The absence of 511\,keV emission from the remnants of recent Galactic SNe Ia turns out to be consistent with the predictions of our simple model. In the most favourable case of positrons being trapped in the ejecta, the emission starts at quite high values but falls by more than 7 orders of magnitude over the first 50\,yrs and then continues decreasing more gently over the following centuries. Our observations of Tycho, Kepler and SN1006 are definitely not constraining (and are therefore not shown in the plot). The upper-limit on the 511\,keV flux from the youngest Galactic SNR is 4 orders of magnitude above the predicted flux for the trapped scenario, and so G1.9+0.3 does not help constraining the process of positron transport and annihilation in SNe/SNRs either.\\
\indent For the two remnants of core-collapse SNe of our set, we computed specific annihilation lightcurves shown in Fig. \ref{lightcurve_ccSNe} where the contributions from both $^{56}$Ni and $^{44}$Ti positrons were added. In the case of Cas A, our upper-limit is more than 3 orders of magnitude above the highest predicted flux, obtained when positrons are trapped in the ejecta. The situation is similar for SN1987A, although our upper-limit is closer to the theoretical lightcurves. In both cases, the INTEGRAL/SPI observations are not constraining enough to favour one scenario of positron transport over the other.\\
\indent As seen above, the highest annihilation fluxes from SNRs are obtained when the decay positrons are trapped in the expanding ejecta. When the latter are free to escape, the corresponding annihilation emission strongly drops within a few decades after the explosion. In this case, however, their slowing-down and annihilation proceed in the CSM/ISM. From our simulation of decay positron annihilation in the CSM/ISM, we found that escaped positrons can give rise to the following steady 511\,keV flux over the first millenary:
\begin{align}
\label{eq_ISMflux}
F_{511}= & K_{ej} \, K_{esc} \, \left( \frac{d}{1\,\textrm{kpc}}\right)^{-2} \left( \frac{n_{0}}{1\,\textrm{cm}^{-3}} \right) \left( \frac{Y_{56}}{0.1\,\textrm{M}_{\odot}}\right)\\ \notag
& \times \, 10^{-5}\,\textrm{ph}\,\textrm{cm}^{-2}\,\textrm{s}^{-1}
\end{align}
where $d$, $n_0$ and $Y_{56}$ are respectively the distance to the object, the CSM/ISM density and the iron yield. $K_{esc}$ is the escape fraction in \% and $K_{ej}$ is a correcting factor accounting for the effect of positron energy loss due to ejecta crossing, which was found empirically to be about 1 for light ejecta of 1-2\,M$_{\odot}$ and about 2 for massive ejecta of 8-10\,M$_{\odot}$. The formula involves only $^{56}$Ni because it was shown in \ref{simu_results} that $^{44}$Ti positrons escaping in the CSM/ISM lead to an annihilation flux that is more than a factor of 10 lower than the flux arising from escaping $^{56}$Ni positrons (see Fig. \ref{lightcurve_ISM_esca}). The contribution of $^{44}$Ti is therefore neglected.\\
\indent From Eq. \ref{eq_ISMflux} and the data in Table \ref{tab_SNRs}, we can compute the expected CSM/ISM annihilation fluxes and convert our upper-limits on the 511\,keV flux into upper-limits on the escape fractions. We eventually obtain upper-limits on the escape fraction of $\sim$13\% for Cas A, $\sim$12\% for Tycho, $\sim$30\% for Kepler and $\sim$33\% for SN1006. SN1987A and G1.9+0.3 are too far away to be constraining. In the case of substantial spatial diffusion of escaped positrons in the CSM/ISM, the upper-limits on the escape fraction in Tycho and SN1006 would be increased by a factor of $\sim$2 (but in this case, our estimate for the surrounding density may not be valid anymore).\\
\indent The value obtained for SN1006 is higher than the 7.5\% estimate from \citet{Kalemci:2006} but was derived from a much finer model and should therefore be considered as more reliable. Our most constraining upper limits of 12 and 13\% on the escape fraction of $^{56}$Ni positrons in Tycho and Cas A, respectively, are consistent with the maximum values obtained from modelling by \citet{Chan:1993} and \citet{Milne:1999} and reported in Sect. \ref{pastworks}. In addition, \citet{Higdon:2009} showed that the steady-state Galactic positron production rate of $\sim2.4 \times 10^{43}$\,e$^{+}$\,s$^{-1}$ and all other properties of the 511\,keV INTEGRAL/SPI observations can be entirely explained by a nucleosynthesis origin if the $^{56}$Ni positrons from SNe Ia have an escape fraction of 5 $\pm$2\%. In this case, $^{56}$Ni would provide about two thirds of the Galactic positrons, the remainder coming from $^{44}$Ti (20\%) and $^{26}$Al (12\%) for which escape from SNe/SNRs is not a problem, and these $\sim$MeV positrons can propagate over fairly large distances in the Galaxy before they annihilate. An escape fraction of 5\% for $^{56}$Ni positrons actually corresponds to the value obtained by \citet{Chan:1993} for the W7 deflagration model of \citet{Nomoto:1984} under the assumption of no ejecta mixing and a combed out magnetic field. It is also consistent with the 5.5\% obtained by \citet{Milne:1999} for the same W7 model, a radial or weak magnetic field, and a 1\% ionisation of the ejecta. Our 12\% upper-limit on the escape fraction of $^{56}$Ni positrons for the Tycho SNIa remnant is therefore consistent with these estimates, and therefore with a nucleosynthesis origin of the Galactic positrons that annihilate at 511\,keV.\\
\indent We emphasise that the above results are associated with the assumptions of a uniform neutral ejecta homologously expanding in a uniform neutral medium. More realistically, supernova ejecta rather follow exponential or power-law density distributions \citep[see for instance][]{Marcaide:1997,Dwarkadas:1998,Truelove:1999} and their expansion is accompanied by the formation of shocks that heat and compress the outermost ejecta and the nearby CSM/ISM. The X-ray radiation from these shocked layers and the cosmic-rays accelerated at the shock fronts may ionise the freely-expanding ejecta to a certain degree. If most of the remnant is ionised, the decay positrons would experience a more efficient slowing-down (by Coulomb losses), but their annihilation timescale may also be longer (if positronium formation in-flight cannot efficiently take place). The shocks and shocked layers are also places of magnetic field amplification and turbulence where the positrons escaping from the freely-expanding ejecta may be trapped. How these effects would modify the predicted signals from positrons annihilating in the SN/SNR ejecta is currently unexplored. For positrons annihilating in the CSM/ISM, the main uncertainty is the ionisation state of the medium. In case of strong ionisation out to large distances, the annihilation timescale in a typical 1\,cm$^{-3}$ medium exceeds the ages of the young SNRs considered in this work. In addition, as mentioned above, a fraction of the escaping positrons can be trapped near the shock fronts and may be reaccelerated there (especially since they are already in the supra-thermal domain). Such effects would allow higher escape fractions than the above values, without conflicting with the INTEGRAL/SPI observations.

\section{Conclusion}
\label{conclu}

\indent We simulated the 511\,keV lightcurves resulting from the annihilation of the decay positrons of $^{56}$Ni and $^{44}$Ti in SNe/SNRs and their surrounding CSM/ISM. We used a simple model of uniform ejecta in homologous expansion and a uniform density medium. Two extreme scenarios of positron transport were considered: either the positron is free to escape the ejecta along radial trajectories, or it is completely trapped in the ejecta.\\
\indent We computed specific 511\,keV lightcurves for Cas A, Tycho, Kepler, SN1006, G1.9+0.3 and SN1987A, and compared these to the upper-limits derived from INTEGRAL/SPI observations. The predicted 511\,keV signals from positrons annihilating in the SN/SNR ejecta are in all cases below the sensitivity of the SPI instrument by several orders of magnitude and therefore do not allow constraining the transport of positrons. From the predicted 511\,keV signals for positrons escaping the ejecta and annihilating in the CSM/ISM, we obtained upper-limits on the $^{56}$Ni positron escape fraction of $\sim$13\% for Cas A, $\sim$12\% for Tycho, $\sim$30\% for Kepler and $\sim$33\% for SN1006. The 12\% upper limit on the $^{56}$Ni positron escape fraction in a SNIa is consistent with the recently estimated value of 5 $\pm$2\% required for a nucleosynthesis origin of the positrons that give rise to the diffuse Galactic emission at 511\,keV.

\begin{acknowledgement}
The SPI project has been completed under the responsibility and leadership of CNES. We are grateful to ASI, CEA, CNES, DLR, ESA, INTA, NASA and OSTC for their support.
\end{acknowledgement}

\bibliographystyle{aa}
\bibliography{/Users/pierrickmartin/Documents/MyPapers/biblio/SNobservations,/Users/pierrickmartin/Documents/MyPapers/biblio/SNmodels,/Users/pierrickmartin/Documents/MyPapers/biblio/SPI,/Users/pierrickmartin/Documents/MyPapers/biblio/26Al&60Fe,/Users/pierrickmartin/Documents/MyPapers/biblio/SNRobservations,/Users/pierrickmartin/Documents/MyPapers/biblio/SNRmodels,/Users/pierrickmartin/Documents/MyPapers/biblio/GalaxyObservations,/Users/pierrickmartin/Documents/MyPapers/biblio/44Ti,/Users/pierrickmartin/Documents/MyPapers/biblio/Positron,/Users/pierrickmartin/Documents/MyPapers/biblio/CassiopeeA,/Users/pierrickmartin/Documents/MyPapers/biblio/CosmicRayTransport,/Users/pierrickmartin/Documents/MyPapers/biblio/Cygnus&CygOB2}

\end{document}